\documentclass{article}
\usepackage{fullpage}

\usepackage{amsthm}
\usepackage{amsmath}
\usepackage{amssymb}
\providecommand{\abs}[1]{\left|#1\right|}

\usepackage{graphicx}
\usepackage{epstopdf}
\usepackage{subcaption}

\usepackage{authblk}

\usepackage{mathtools} 
\newtheorem*{remark*}{Remark}
\newtheorem*{theorem*}{Theorem}

\newcounter{saveeqn}
\newcommand{\alpheqn}{\setcounter{saveeqn}{\value{equation}}
\stepcounter{saveeqn}\setcounter{equation}{0}
\renewcommand{\theequation}{\mbox{\arabic{saveeqn}\alph{equation}}}}

\newcommand{\resetalpheqn}{\setcounter{equation}{\value{saveeqn}}
\renewcommand{\theequation}{\arabic{equation}}}

\providecommand{\keywords}[1]{\textbf{\textit{Keywords:}} #1}

\begin{document}
\pagestyle{plain}
\title{Numerical inverse scattering for the sine-Gordon equation}

\author[$\dagger$]{Bernard Deconinck}
\author[$*$]{Thomas Trogdon}
\author[$\dagger$]{Xin Yang}
\affil[$\dagger$]{Department of Applied Mathematics, University of Washington, Seattle}
\affil[$*$]{Department of Mathematics, University of California, Irvine}

\maketitle

\begin{abstract}
We implement the numerical inverse scattering transform (NIST) for the sine-Gordon equation in laboratory coordinates on the real line using the method developed by Trogdon, Olver and Deconinck \cite{kdv}. The NIST allows one to compute the solution at any $x$ and $t$ without having spatial discretization or time-stepping. The numerical implementation is fully spectrally accurate. With the help of the method of nonlinear steepest descent, the NIST is demonstrated to be uniformly accurate. 
\end{abstract}
\keywords{sine-Gordon equation; numerical inverse scattering transform; Riemann-Hilbert problem; nonlinear steepest descent}
\section{Introduction}
We consider the sine-Gordon (SG) equation in laboratory coordinates on the real line,
\begin{align}
u_{tt}-u_{xx}+\sin(u)=0,\,\,\, x\in \mathbb{R},\,\,t\geq 0.
 \label{sg}
\end{align}
The SG equation is a nonlinear partial differential equation which appears in differential geometry and various applications such as superconductivity and Josephson junctions \cite{barone}. Many numerical methods have been developed to solve the SG equation \cite{guo,jiang,shukla}. Using these methods, or other more traditional but less specialized methods, it is hard to obtain the solution accurately, especially for long time \cite{ablowitz1,ablowitz2}. In addition, working on an unbounded domain requires special treatment since most traditional methods require domain truncation ~\cite{zheng}.

Ablowitz, Kaup, Newell and Segur \cite{AKNS1} were the first to show that the Cauchy problem for the SG equation written in light-cone coordinates,
\[
u_{xt}=\sin(u),
\]
is integrable and can be solved by the inverse scattering transform (IST) method. Kaup \cite{kaup} demonstrated that (\ref{sg}) is solvable by the IST method. This is important as (\ref{sg}) is the relevant form of the SG equation for most applications.  In 2012, Trogdon, Olver and Deconinck implemented the numerical inverse scattering transform (NIST) for the Korteweg-de Vries (KdV) and modified Korteweg-de Vries (mKdV) equations \cite{kdv}. The NIST is applied successfully to other integrable systems such as the focusing and defocusing nonlinear Schr\"{o}dinger (NLS) equations \cite{nls} and the Toda lattice \cite{bilmantrogdon}. The NIST makes no domain approximation, does not require time-stepping and is uniformly accurate. As such, it provides a benchmark for other numerical methods \cite{bilmantrogdon2}. We want to solve (\ref{sg}) for $x, t\in \mathbb{R}, t\geq0$ using the NIST. We assume that the initial values, $\sin(u(x,0))$ and $\sin(u_t(x,0))$ are in $S_{\delta}(\mathbb{R})$, {\em i.e.}, Schwartz-class functions on the real line $S(\mathbb{R})$ with exponential decay:
\begin{align}
S_{\delta}(\mathbb{R})=\left\{ f\in S(\mathbb{R}): \abs{f(x)}e^{\delta \abs{x}}\rightarrow 0, \, as \abs{x}\rightarrow \infty \mbox{ for } \delta>0\right\}.
\label{schwartz}
\end{align}
The decay and regularity requirements are mainly for numerical convenience. The global well-posedness theory of the SG equation only assumes initial values in $L^p(\mathbb{R})$ \cite{buckinghammiller}.

\begin{figure}
  \centering
  \begin{subfigure}{0.4\textwidth}
  \center
  \includegraphics[width=0.8\textwidth]{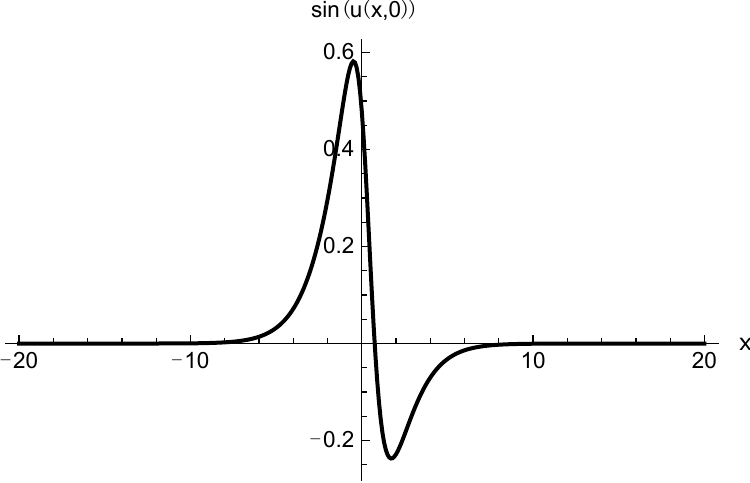}
  \caption{The initial value $\sin(u(x,0))$ is a two-soliton solution with a sech$^2$ perturbation. The error is on the order of $10^{-10}$. }
  \end{subfigure}
    \hspace{1cm}
    \begin{subfigure}{0.4\textwidth}
  \center
  \includegraphics[width=0.8\textwidth]{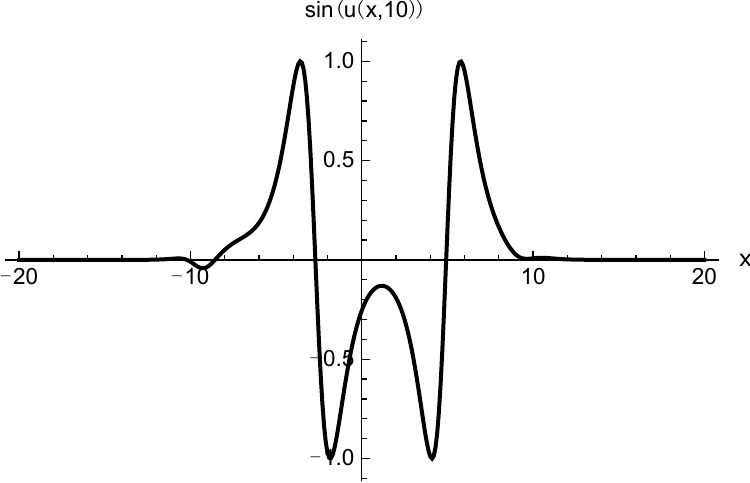}
  \caption{Numerical solution $\sin(u(x,10))$. Two solitons separate from each other. Dispersive effects starts to appear near $x=-10$.}
  \end{subfigure}\\
  \begin{subfigure}{0.4\textwidth}
  \center
  \includegraphics[width=0.8\textwidth]{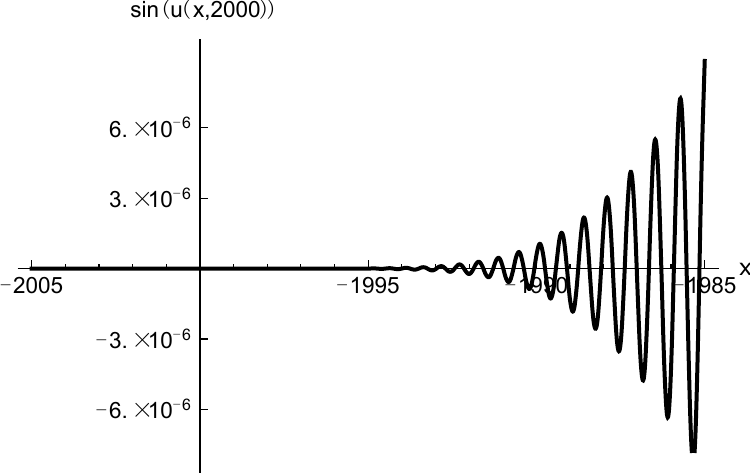}
  \caption{The exponentially growing oscillatory solution near $x=-2000$ at $t=2000$.}
  \end{subfigure}
      \hspace{1cm}
    \begin{subfigure}{0.4\textwidth}
  \center
  \includegraphics[width=0.88\textwidth]{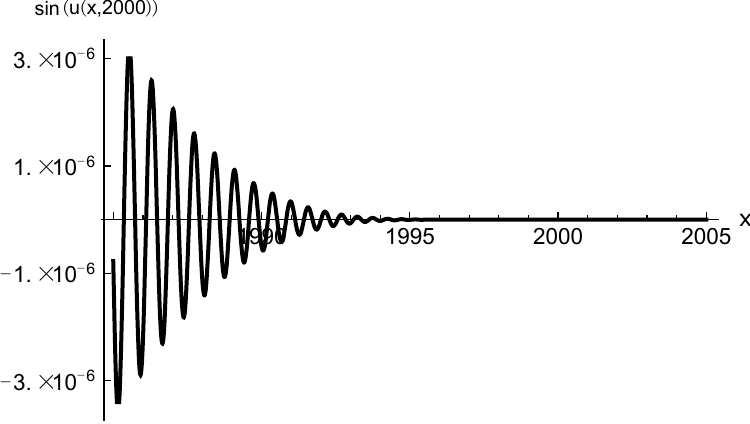}
  \caption{The exponentially decaying oscillatory solution near $x=2000$ at $t=2000$.}
  \end{subfigure}\\
  \begin{subfigure}{0.4\textwidth}
  \center
  \includegraphics[width=0.8\textwidth]{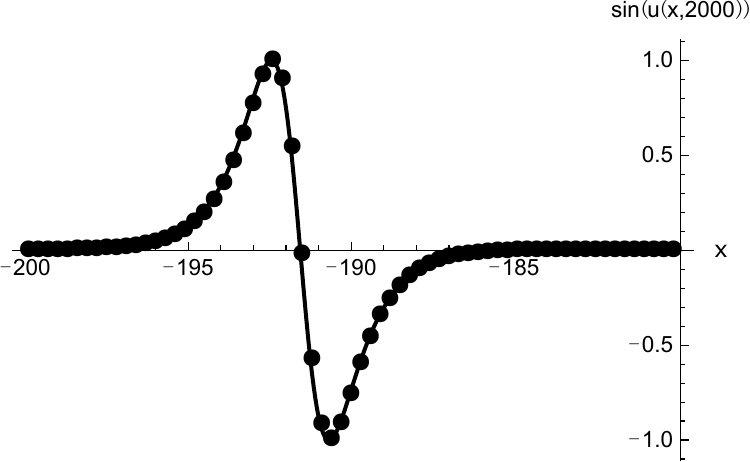}
  \caption{The soliton near $x=-190$ at $t=2000$. (Dots) numerical
   solution, and (Solid) shifted unperturbed exact solution.}
  \end{subfigure}
      \hspace{1cm}
    \begin{subfigure}{0.4\textwidth}
    \center
  \includegraphics[width=0.72\textwidth]{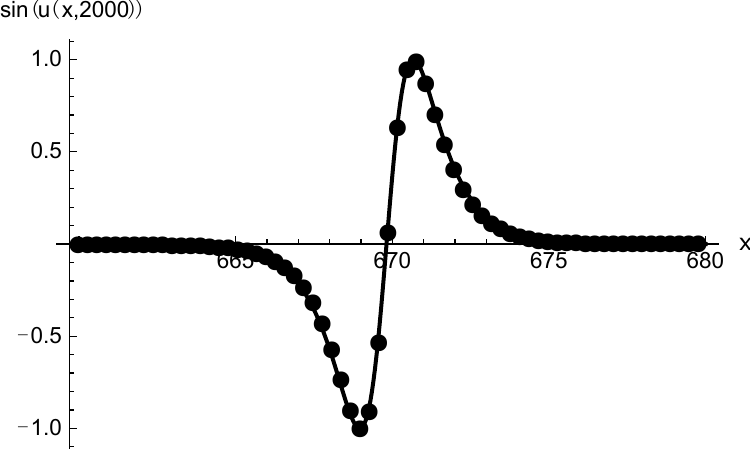}
  \caption{The soliton near $x=670$ at $t=2000$.\\ (Dots) numerical
   solution, and (Solid) shifted unperturbed exact solution.}
  \end{subfigure}\\
  \begin{subfigure}{\textwidth}
  \center
  \includegraphics[width=0.9\textwidth]{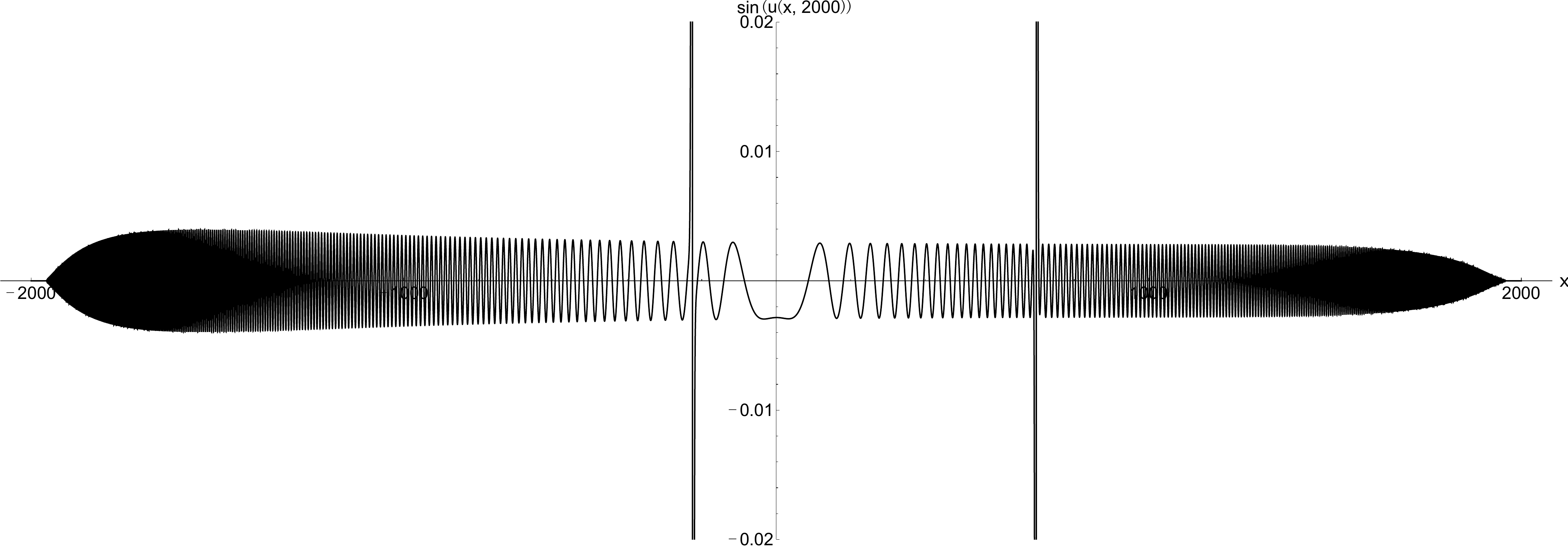}
  \caption{The numerical solution $\sin(u(x,2000))$ for $x$ from $-2000$ to $2000$. The two spikes are the solitons with amplitude $1$ while the amplitude of the dispersive waves is about $0.005$.}
  \end{subfigure}
  \caption{The numerical solution $\sin(u(x,t))$ of (\ref{sg}). The initial value is $u(x,0)=v(x,0)+0.5\mbox{sech}^2(x)$, $u_t(x,0)=v_t(x,0)$. The two-soliton solution $v(x,t)$ is generated by the consistency condition (\ref{bt}) using two one-soliton solutions (\ref{1soliton}) with $k_1=\sqrt{3/5}$, $k_2=1$ and a zero solution. }\label{introp}
\end{figure}

\begin{figure}
  \centering
  \begin{subfigure}{0.4\textwidth}
  \center
  \includegraphics[width=0.8\textwidth]{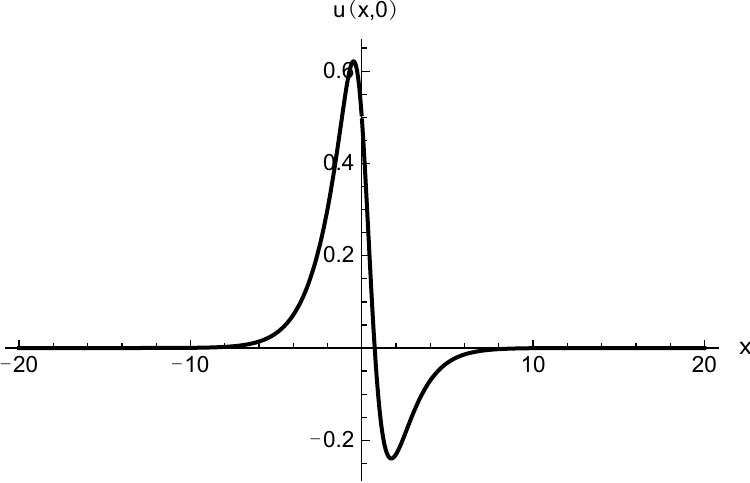}
  \caption{The initial value $u(x,0)$ is a two-soliton solution with a sech$^2$ perturbation. The error is on the order of $10^{-10}$. }
  \end{subfigure}
    \hspace{1cm}
  \begin{subfigure}{0.4\textwidth}
  \center
  \includegraphics[width=0.8\textwidth]{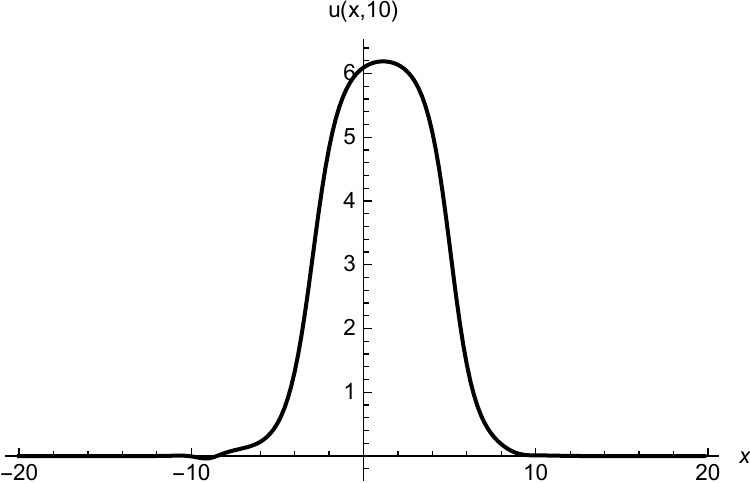}
  \caption{The numerical solution $u(x,10)$.\\ Two solitons separate from each other. Dispersive effects starts to appear near $x=-10$. }
  \end{subfigure}
  \begin{subfigure}{0.4\textwidth}
  \center
  \includegraphics[width=0.8\textwidth]{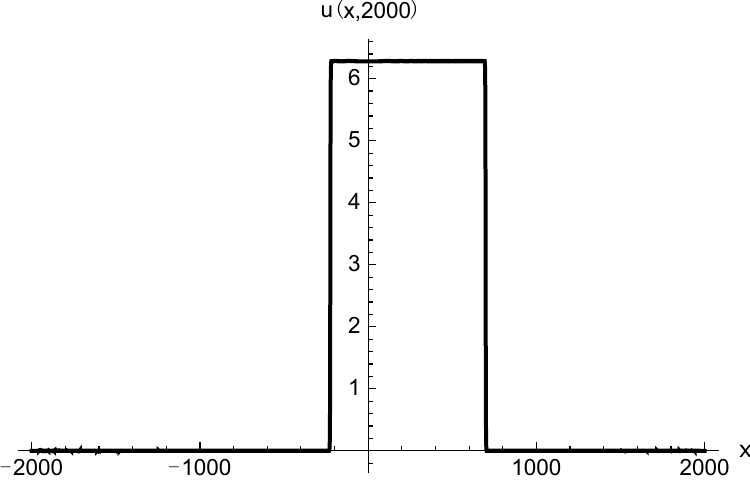}
  \caption{The numerical solution $u(x,2000)$ for $x$ from $-2000$ to $2000$. The amplitude of the dispersive waves is on the order of $0.005$.}
  \end{subfigure}
      \hspace{1cm}
  \begin{subfigure}{0.4\textwidth}
  \center
  \includegraphics[width=0.8\textwidth]{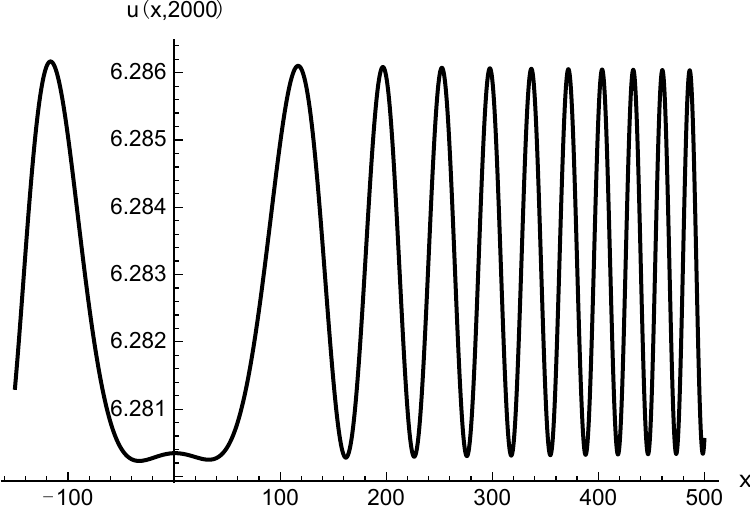}
  \caption{The dispersive waves between the two solitons.}
  \end{subfigure}
  \\
  \begin{subfigure}{0.4\textwidth}
  \center
  \includegraphics[width=0.8\textwidth]{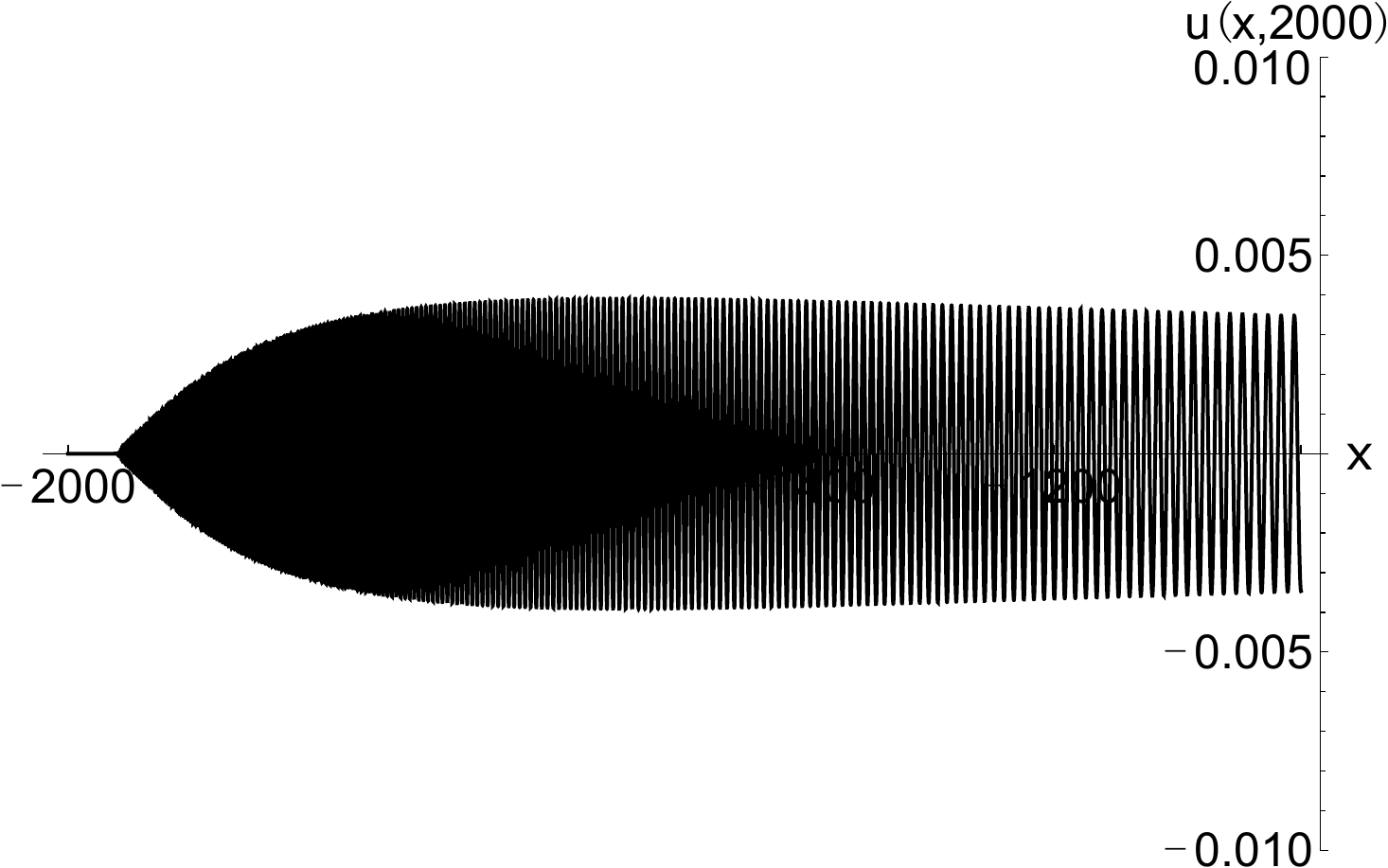}
  \caption{The dispersive waves in $x\in [-2000,-1000]$ at $t=2000$.}
  \end{subfigure}
      \hspace{1cm}
    \begin{subfigure}{0.4\textwidth}
  \center
  \includegraphics[width=0.88\textwidth]{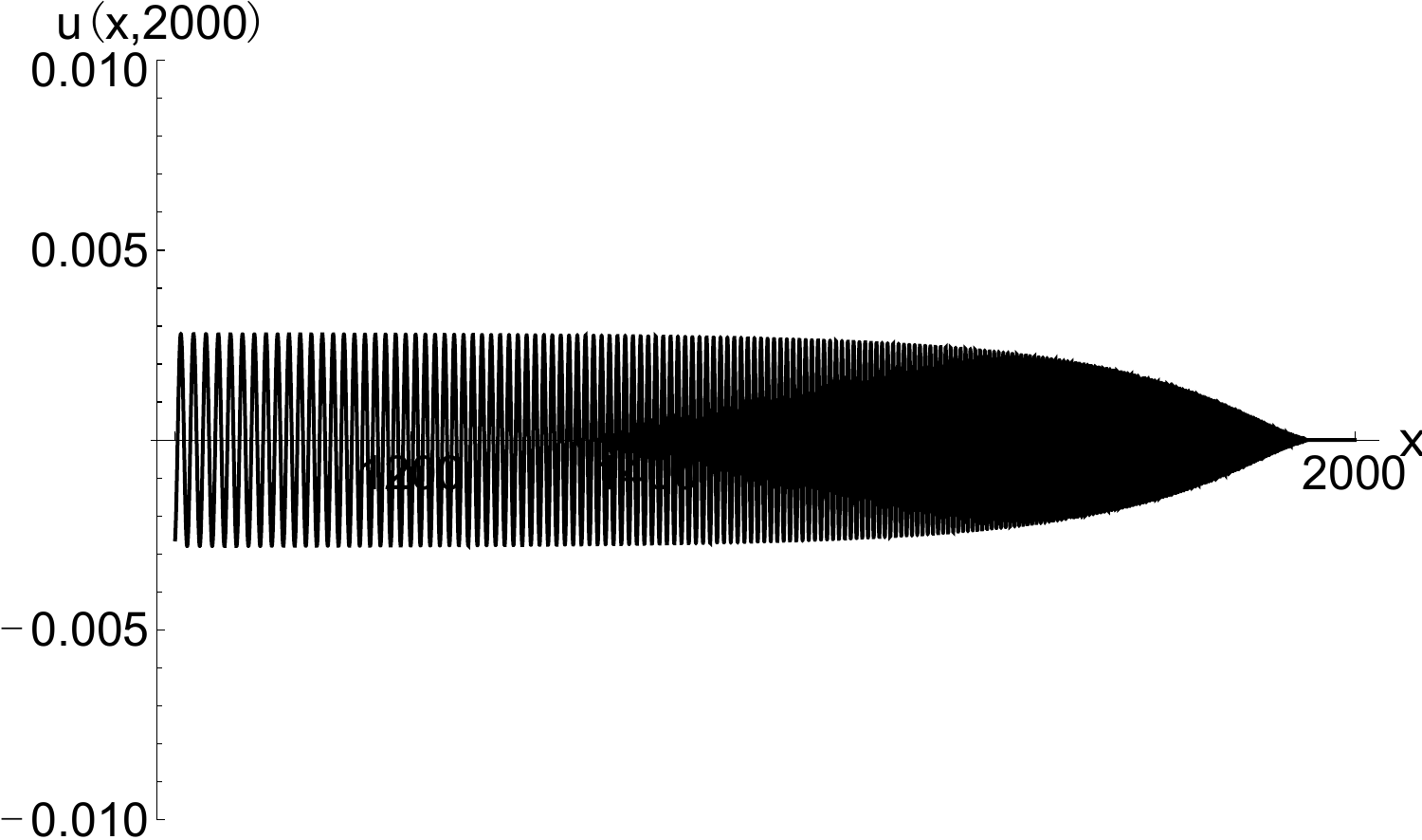}
  \caption{The dispersive waves in $x\in [1000,2000]$ at $t=2000$.}
  \end{subfigure}\\
  \caption{The numerical solution $u(x,t)$ of (\ref{sg}). The initial values are the same as in Figure \ref{introp}. This demonstrates that the high oscillations are not due to a rapidly growing argument of sine but are inherent to the solution itself. 
   }\label{intropu}
\end{figure}

We illustrate the complex structure and highly oscillatory behaviour of the solution. Figure \ref{introp} shows the numerical solution of (\ref{sg}) at $t=10$ and at $t=2000$ using our numerical inverse scattering transform (NIST) with the initial values in Figure \ref{introp}(a) given by a perturbed two-soliton solution. In Figure \ref{introp}(b), at $t=10$, the dispersive wave is apparent with an approximate amplitude $0.05$ near $x=-10$. At $t=2000$, the amplitude of the dispersive wave decays to $0.005$ and is more oscillatory. Near $x=\pm 2000$ at $t=2000$, the solution decays exponentially fast to $0$, see Figure \ref{introp}(c,d). The uniform accuracy of the NIST guarantees that the numerical solution does not lose accuracy for larger time and in Figure \ref{introp}(e,f) we see that the profiles of the two solitons are preserved. In Figures \ref{introp} and \ref{intropu}, both $\sin(u)$ and $u$ are shown respectively to demonstrate that the oscillations of the $\sin(u)$ are not due to large growth of $\abs{u}$.     

\section{Integrability and Riemann-Hilbert problems}
Before we construct the NIST, we need the details of the IST for (\ref{sg}). Most of the results are from \cite{buckinghammiller,cheng,kaup}. For consistency, we present the method using the style and notation of \cite{tom}.
The SG equation is completely integrable with Lax pair \cite{kaup}:
\alpheqn
\begin{align}
                  \psi_x&=X(z,u,u_x,u_t) \psi \label{lpx},\\
                  \psi_t&=T(z,u,u_x,u_t) \psi \label{lpt},
\end{align}
\resetalpheqn
where
\[
X(z,u,u_x,u_t)=\left(
    \begin{array}{cc}
      -\displaystyle\frac{iz}{4} & 0 \\
      0 & \displaystyle\frac{iz}{4} \\
    \end{array}
  \right)
  +\left(
            \begin{array}{cc}
              \displaystyle\frac{i\cos(u)}{4z} &  \displaystyle\frac{i\sin(u)}{4z} \\  [1em]
               \displaystyle\frac{i\sin(u)}{4z} &  -\displaystyle\frac{i\cos(u)}{4z} \\
            \end{array}
          \right)
  +\left(
     \begin{array}{cc}
       0 & -\displaystyle\frac{u_x+u_t}{4} \\
       \displaystyle\frac{u_x+u_t}{4} & 0 \\
     \end{array}
   \right),
\]
and
\[
T(z,u,u_x,u_t)=\left(
    \begin{array}{cc}
      -\displaystyle\frac{iz}{4} & 0 \\
      0 & \displaystyle\frac{iz}{4} \\
    \end{array}
  \right)
  -\left(
            \begin{array}{cc}
              \displaystyle\frac{i\cos(u)}{4z} &  \displaystyle\frac{i\sin(u)}{4z} \\  [1em]
               \displaystyle\frac{i\sin(u)}{4z} &  -\displaystyle\frac{i\cos(u)}{4z} \\
            \end{array}
          \right)
  +\left(
     \begin{array}{cc}
       0 & -\displaystyle\frac{u_x+u_t}{4} \\
       \displaystyle\frac{u_x+u_t}{4} & 0 \\
     \end{array}
   \right).
\]
The existence of a joint matrix solution $\psi$ satisfying both equations (\ref{lpx},b) requires the compatibility condition $\psi_{xt}=\psi_{tx}$, which is an equivalent representation of the SG equation (\ref{sg}) \cite{kaup}. In this new representation, the solution to the SG equation (\ref{sg}) can be obtained by solving linear equations. In the IST method, (\ref{lpx}) determines the scattering data defined in the following section and (\ref{lpt}) determines the time evolution of this scattering data.
\subsection{Direct scattering}
The process of finding the scattering data from (\ref{lpx}) is called direct scattering. Because of the compatibility condition, (\ref{lpx}) can be solved at any value of the parameter $t$. To obtain the scattering data, we define two matrix solutions to (\ref{lpx}) by their corresponding asymptotic behavior.
\begin{align*}
\psi^-(x,t,z) &\sim \left(
           \begin{array}{cc}
             e^{-ix(z-1/z)/4} & 0 \\
             0 & -e^{ix(z-1/z)/4} \\
           \end{array}
         \right), \mbox{ as } x\rightarrow -\infty,\\
\psi^+(x,t,z) &\sim \left(
           \begin{array}{cc}
             e^{-ix(z-1/z)/4} & 0 \\
             0 & e^{ix(z-1/z)/4} \\
           \end{array}
         \right), \mbox{  as } x\rightarrow \infty.
\end{align*}
Since these two solutions are linearly dependent, there exists a scattering matrix $S(z,t)$, independent of $x$, relating $\psi^-(x,t,z)$ and $\psi^+(x,t,z)$:
\begin{align}
\psi^+(x,t,z)=\psi^-(x,t,z)S(z,t).
\label{scatteringm}
\end{align}
The matrix $S(z,t)$ plays the role of the Fourier transform in linear problems \cite{AKNS1,kaup} and can be used to recover the function $u(x,t)$ by inverse scattering. In practice, different forms of the scattering matrix are used.
Let
\[
S(z,t)=\left( \begin{array}{cc}
                                                                            a(z,t) & B(z,t) \\
                                                                            b(z,t) & A(z,t) \\
                                                                             \end{array}
                                                                           \right),
                                                                           \]
which for real initial values has the symmetry \cite{kaup},
\begin{align}
a(z,t)=-\overline{A(\bar{z},t)},\quad b(z,t)=\overline{B(\bar{z},t)},
\label{symmetry1}
\end{align}
and
\begin{align}
a(z,t)=-A(-z,t),\quad b(z,t)=B(-z,t).
\label{symmetry2}
\end{align}
Moreover, with the Schwartz-class initial values, $a(z,t)$ is analytic in the upper-half $z$-plane and $a(z,t)$, $b(z,t)$ are bounded on the real $z$-axis \cite{kaup}. The exponential decay of the initial values in $S_{\delta}(\mathbb{R})$ allows $a(z,t)$ and $b(z,t)$ to be analytically extended to regions defined by $\mathcal{D}_{\delta}:=\{\abs{\mbox{Re}(i(z-1/z)/4)}<\delta/2\}$ which looks like a strip pinched at the origin. The boundary of $\mathcal{D}_{\delta}$ in the first quadrant of the complex $z-$plane is a level set of $\abs{\mbox{Re}(i(z-1/z)/4)}$ shown in Figure \ref{conditionnumber} and is symmetric with respect to the real and imaginary axes. For some $u(x,0)$ and $u_t(x,0)$, there may exists values $z=\kappa_j$, $\mbox{\emph{Im}}(\kappa_j)>0$ such that $a(\kappa_j,0)=0$. These correspond to bound states. The number of bound states is finite for initial values in $S_{\delta}$. As in \cite{cheng,kaup}, we assume that the zeros are simple and not real. This is guaranteed in the case of compactly supported initial values, is true in many other cases \cite{buckinghammiller,huanglenells} and in all the numerical examples we have computed. (For instance, we get non-zero or real zeros of $a(z)$ if ($\gamma(\mu)-1)/2\epsilon\in \mathbb{N}^+$ in (\ref{exactrho}) \cite{buckinghammiller}). \\
Define the reflection coefficient
\[\rho(z,t):=\frac{b(z,t)}{a(z,t)}.\]
For $z\in \mathbb{R}$, and fixed $t$, $\rho(z,t)\in S(\mathbb{R})$ is a Schwartz-class function and $\abs{\rho(z,t)}\rightarrow 0$ faster than any power of $z$ as $\abs{z}\rightarrow 0$ \cite{cheng}. It is important to note that $b(z,t)$ is only defined on the real line and may or may not have an analytic continuation off the real axis. However, $b(\kappa_j,t)$ is defined at the zeros of $a(z,t)$ in the upper-half $z$-plane as a proportionality coefficient determined by solving (\ref{scatteringm}) directly \cite{tom}. For instance, we can have $b(z,t)=0$ for real $z$ but $b(\kappa_j,t)\neq 0$ which is the corresponding scattering data for a pure soliton solution. With exponentially decaying initial values, $\rho(z,t)$ is guaranteed to have an analytic continuation near the real $z$-axis except at $z=0$ and $z=\infty$. Moreover, if the initial values are compactly supported, $a(z,t)$ and $b(z,t)$ are analytic everywhere except at $z=0$ and $z=\infty$ \cite{kaup}. At the zeros of $a(z,t)$, $b(\kappa_j,t)$ is the proportionality constant between two fundamental solutions that are exponentially decaying in different directions,
\begin{align}
\psi_2^+(x,t,\kappa_j)=\psi_1^-(x,t,\kappa_j)b(\kappa_j,t),
\label{l2eigen}
\end{align}
where subscripts refer to columns. This implies that $\psi_2^+(x,t,\kappa_j),\psi_1^-(x,t,\kappa_j)$ are eigenfunctions of the Lax operator in (\ref{lpx}). For these values of $z$, considering only simple zeros, the norming constants are defined as
\[
C_j(t)=\frac{b(\kappa_j,t)}{a'(\kappa_j,t)}.
\]
The collection
\[
\mathbb{S}=\{\rho(z,t),\{(\kappa_j,C_j(t))\}_{j=1}^{n}\},
\]
defines the scattering data.
\subsection{Time evolution of the scattering data}
The scattering data $\mathbb{S}$ has simple time dynamics. It is independent of $x$ and its $t$ dependence is explicit. If we choose $x\rightarrow \infty$, (\ref{lpt}) is diagonalized with constant coefficients depending only on $z$. Therefore we can write down the time evolution of the scattering data by plugging (\ref{lpt}) into (\ref{scatteringm}), leading to
\begin{align}
a(z,t)&=a(z,0),\\
b(z,t)&=\exp\left(\frac{it}{4}\left(z+\frac{1}{z}\right)\right)b(z,0).
\label{timeevolution}
\end{align}
It follows that $\{\kappa_j\}$, the zeros of $a(z,t)$, are fixed as time evolves which is an essential component of IST theory. For convenience we suppress the time dependence in $a$, $b$ and $\rho$ if $t=0$.
\subsection{Inverse scattering}
The process of recovering the solution to the SG equation (\ref{sg}) from the scattering data $\mathbb{S}$ is called inverse scattering. We shall perform inverse scattering using Riemann-Hilbert problems (RHPs) \cite{tom}.
Let
\[
m(x,t,z)=\psi(x,t,z) \left(
           \begin{array}{cc}
             e^{ix(z-1/z)/4} & 0 \\
             0 & e^{-ix(z-1/z)/4} \\
           \end{array}
         \right).
\]
Then $m(x,t,z)$ satisfies the ordinary differential equation
\begin{align}
m_x(x,t,z)=[J(z),m(x,t,z)]+Q(x,t,z)m(x,t,z),
\label{ode1}
\end{align}
where
\begin{align}
J(z)=-\frac{i}{4}\left(z-\frac{1}{z}\right)\left(
                         \begin{array}{rr}
                           1 & 0 \\
                           0 & -1 \\
                         \end{array}
                       \right),
\end{align}
\begin{align}
Q(x,t,z)=\left(
         \begin{array}{cc}
           \displaystyle\frac{i}{4z}(\cos(u(x,t))-1) & \displaystyle\frac{i}{4z}\sin(u(x,t))-\frac{1}{4}(u_x(x,t)+u_t(x,t)) \\
           \displaystyle\frac{i}{4z}\sin(u(x,t))+\frac{1}{4}(u_x(x,t)+u_t(x,t)) & -\displaystyle\frac{i}{4z}(\cos(u(x,t))-1) \\
         \end{array}
       \right),
       \label{Q}
\end{align}
and $[\cdot,\cdot]$ is the matrix commutator.
Let $m^+=(m_1^+,m_2^+)$ and $m^-=(m_1^-,m_2^-)$ be the solutions corresponding to $\psi^+$ and $\psi^-$. They satisfy the asymptotic conditions
\[
\lim_{x\rightarrow \infty} m^+(x,t,z)=\left(
              \begin{array}{cc}
                1 & 0 \\
                0 & 1 \\
              \end{array}
            \right), \lim_{x\rightarrow -\infty} \,m^-(x,t,z)=\left(
              \begin{array}{rr}
                1 & 0 \\
                0 & -1 \\
              \end{array}
            \right).
\]
Two new matrices $\hat{\Phi}^+$ and $\hat{\Phi}^-$ are defined by rearranging columns of $m^+$ and $m^-$,
\[
\hat{\Phi}^+(z,x,t)=(m_1^-(x,t,z),m_2^+(x,t,z)),
\]
\[
\hat{\Phi}^-(z,x,t)=(m_1^+(x,t,z),m_2^-(x,t,z)).
\]
Let
\[
\Phi^+(z,x,t)=\hat{\Phi}^+(z,x,t)\left(
         \begin{array}{cc}
           1/a(z,t) & 0 \\
           0 & 1 \\
         \end{array}
       \right),
\]
\[
\Phi^-(z,x,t)=\hat{\Phi}^-(z,x,t)\left(
         \begin{array}{cc}
           1 & 0 \\
           0 & -1/\overline{a(\bar{z},t)} \\  
         \end{array}
       \right).
\]
It has been shown that $\Phi^+$ can be analytically continued to the upper-half $z$-plane, while $\Phi^-$ can be analytically continued to the lower-half $z$- plane if $u_x,u_t$ and $\sin(u/2)$ are integrable \cite{kaup}, which is true if $\sin(u(x,0))$ and $\sin(u_t(x,0))$ are in $S_{\delta}(\mathbb{R})$. Therefore for $z\in \mathbb{R}$, there is a jump condition,
\[
\Phi^+(z,x,t)=\Phi^-(z,x,t)G(z,x,t).
\]
Given the jump function $G(z,x,t)$ on the contour (the real line), the problem of finding $\Phi^+$ analytic in the upper-half $z$-plane and $\Phi^-$ analytic in the lower-half $z$-plane is an (analytic) RHP. The contour is assigned orientation and we use $\Phi^+$ to denote the non-tangential pointwise limit from the left of the contour and $\Phi^-$ for the limit from the right of the contour. After incorporating the zeros of $a(z)$, we arrive at the following (meromorphic) RHP \cite{cheng}.
\begin{theorem*} With the previous definitions of variables, assume that sine of the initial values of~(\ref{sg}), $\sin(u(x,0))$ and $\sin(u_t(x,0))$, are in the space $S_{\delta}(\mathbb{R})$. Assume that $a(z)$ has only simple zeros in the upper-half z-plane. Then there exists a unique function $\Phi(z,x,t)$, $x,t\in \mathbb{R}$, $z\in \mathbb{C}\backslash \mathbb{R}$ continuous up to the real axis satisfying the jump condition:
\begin{align}
\Phi^+(z,x,t)=\Phi^-(z,x,t)G(z,x,t), \,\,\,z\in \mathbb{R}.
\label{rhp}
\end{align}
The jump matrix $G(z,x,t)$ is defined by
\begin{align}
G(z,x,t):=\left(
                        \begin{array}{cc}
                          1+\rho(z)\overline{\rho(\bar{z})} & \overline{\rho(\bar{z})}e^{-\theta(z,x,t)} \\
                          \rho(z)e^{\theta(z,x,t)} & 1 \\
                        \end{array}
                      \right),
\end{align}
where $\rho(z)$ is the reflection coefficient and
\[
\theta(z,x,t)=\frac{i}{2}\left[\left(z-\frac{1}{z}\right)x+\left(z+\frac{1}{z}\right)t\right].
\]
Moreover, $\Phi$ satisfies the asymptotic condition,
\[
\lim_{z\rightarrow \infty}\Phi(z,x,t)=\left(
            \begin{array}{cc}
              1 & 0 \\
              0 & 1 \\
            \end{array}
          \right),
\]
and the residue conditions on zeros on $a(z)$ in the upper-half $z-$plane, $\{\kappa_j|a(\kappa_j)=0,\,\,\mbox{\emph{Im}}(\kappa_j)>0,\,\,j=1,\ldots,N<\infty\}$,
\[
\mbox{\emph{Res}}\{\Phi(z,x,t),z=\kappa_j\}=\lim_{k\rightarrow k_j}\Phi(z,x,t)\left(
                                                       \begin{array}{cc}
                                                         0 & 0 \\
                                                         C_je^{\theta(\kappa_j,x,t)} & 0 \\
                                                       \end{array}
                                                     \right),
\]
\[
\mbox{\emph{Res}}\{\Phi(z,x,t),z=\overline{\kappa_j}\}=\lim_{k\rightarrow \overline{k_j}}\Phi(z,x,t)\left(
                                                       \begin{array}{cc}
                                                         0 & -\overline{C_j}e^{-\theta(\overline{\kappa_j},x,t)} \\
                                                         0 & 0 \\
                                                       \end{array}
                                                     \right).
\]
The solution $\Phi$ is meromorphic in the upper-half and lower-half $z$-plane with its pointwise limit functions on the real line $\Phi^+$ from above and $\Phi^-$ from below that are both continuous.
The corresponding solution to (\ref{sg}) is given by
\begin{align}
u(x,t)=\Phi(0,x,t)\left(
             \begin{array}{rr}
               1 & 0 \\
               0 & -1 \\
             \end{array}
           \right)
\Phi(0,x,t)^{-1}.
\label{recoveru}
\end{align}
\end{theorem*}

For the proof of the theorem, the reader is referred to Theorem 2,3 in \cite{cheng} and \cite{bealscoifman,zhou}.
It should be mentioned that (\ref{recoveru}) is well defined because the reflection coefficient vanishes at $z=0$ implying $\Phi^+(0)=\Phi^-(0)$. In practice (see the next section), the residue conditions are replaced by jump conditions on small circles centered at each pole $\kappa_j$ \cite{kdv,nls}.

\section{Numerical direct scattering}
\subsection{Computing the reflection coefficient}
\begin{figure}
  \centering
  \includegraphics[width=0.49\textwidth]{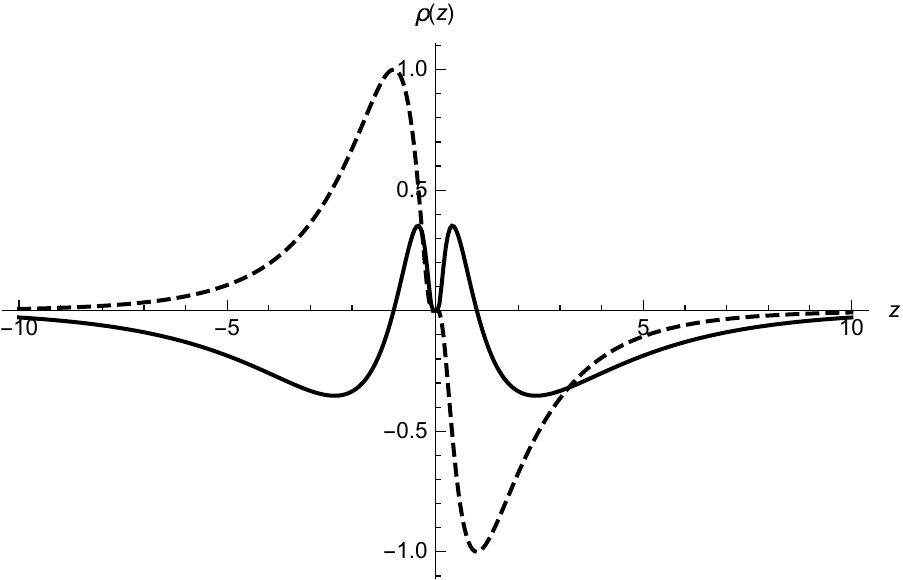}
  \includegraphics[width=0.49\textwidth]{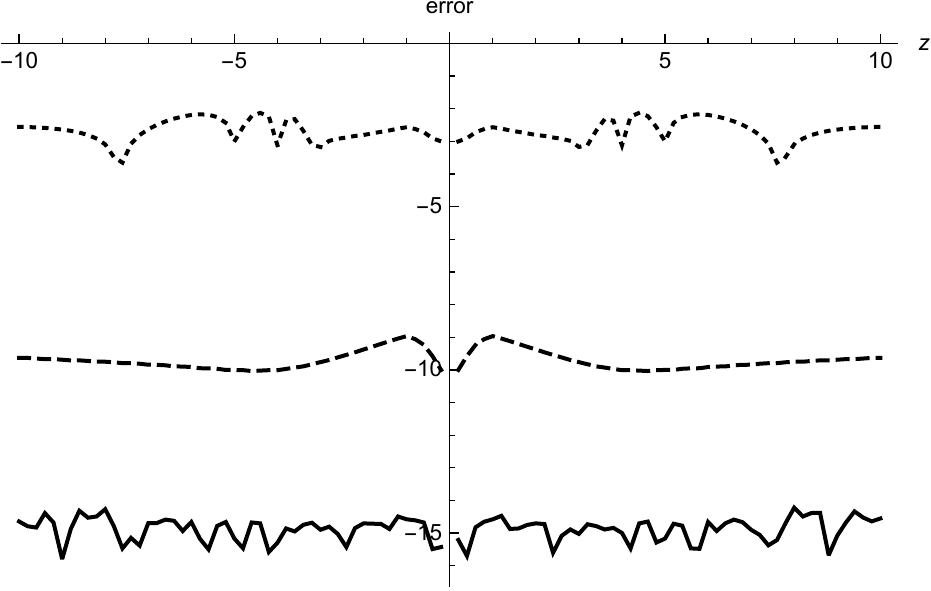}
  \caption{(Left) The exact reflection coefficient $\rho(z)$, $z\in[-10,10]$. Solid: real part. Dashed: imaginary part. (Right) The plot of log of the absolute error for $z\in[-10,10]$. Dotted: 20 collocation points. Dashed: 70 collocation points. Solid: 120 collocation points.}\label{comprho}
\end{figure}

\begin{figure}
  \centering
  \includegraphics[width=0.45\textwidth]{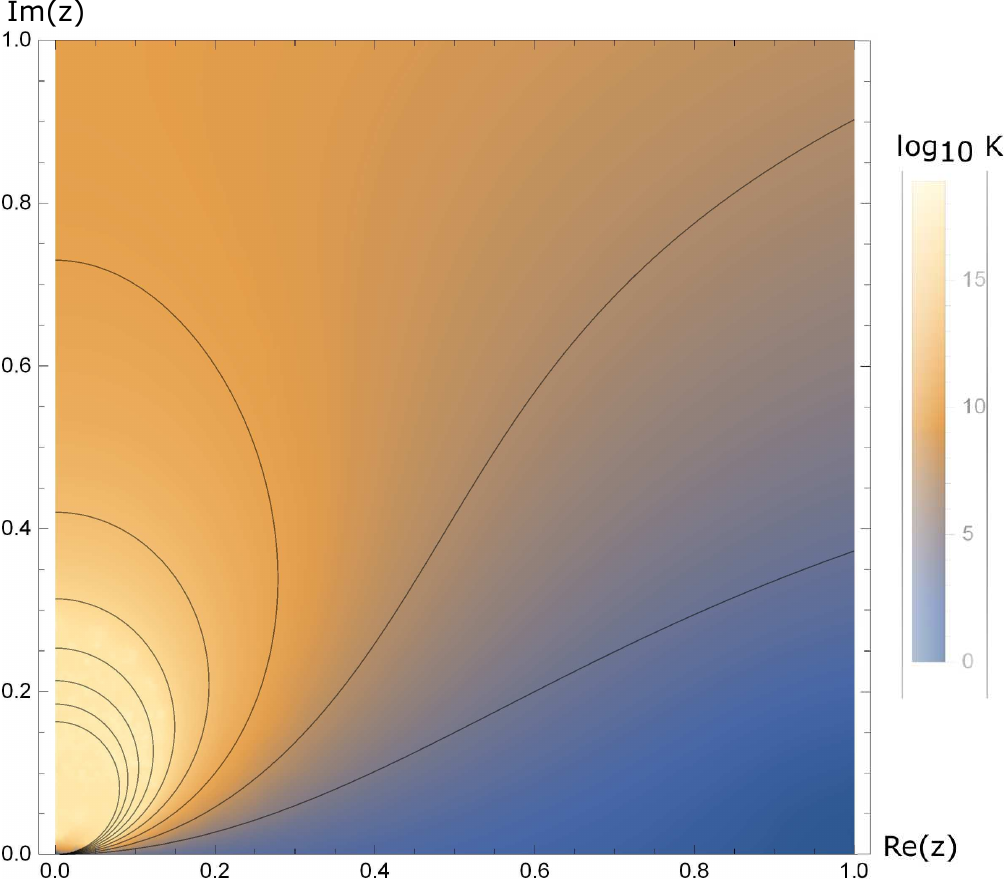}
  \includegraphics[width=0.45\textwidth]{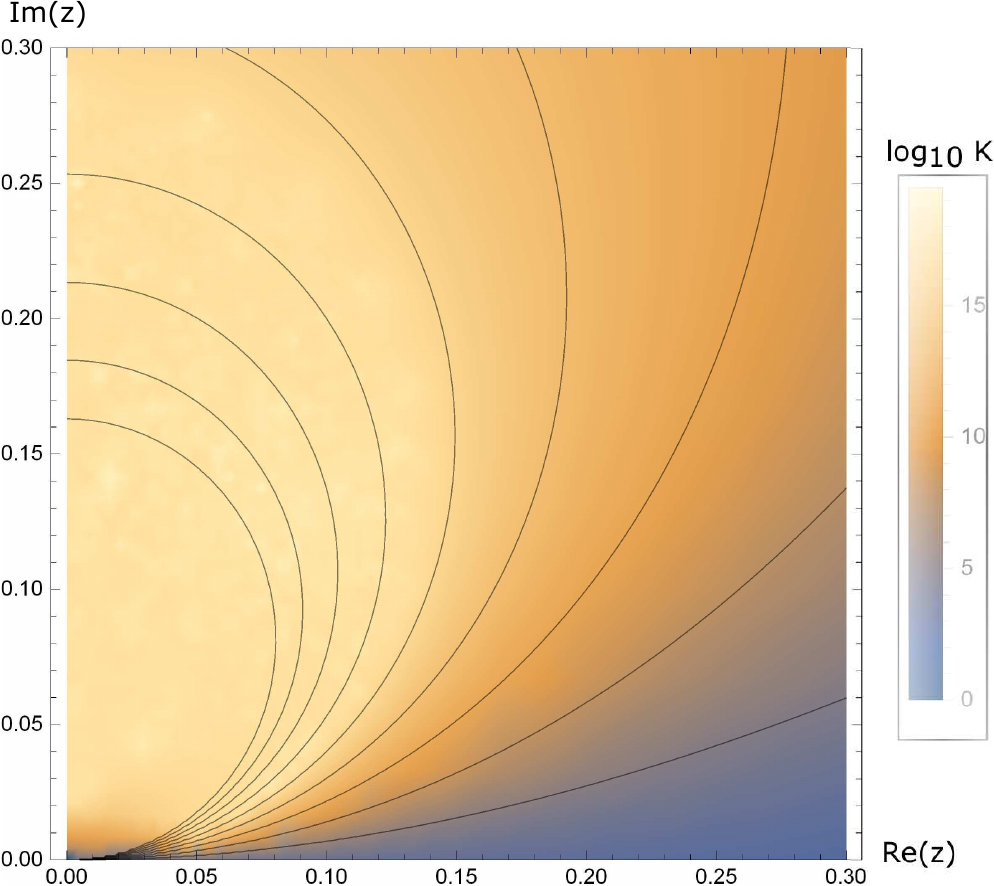}
f  \caption{The condition number K (in $\mbox{log}_{10}$ scale) for the linear system using the Chebyshev collocation method with $120$ collocation points for $0\leq \mbox{Re}(z),\,\mbox{Im}(z) \leq 1$ (left) and for $0\leq \mbox{Re}(z),\,\mbox{Im}(z) \leq 0.3$ (right). The contour lines are the level sets of $\abs{\mbox{Re}(i(z-1/z)/4)}$. These level sets match the condition number except in a small neighbourhood of the origin where the condition number is moderate, i.e., the dark colors extend to the origin in the right panel. In this small neighbourhood we can use straight lines with nonzero slope for the deformations in Section \ref{secnist}.}\label{conditionnumber}
\end{figure}

In order to obtain the scattering data from the initial values, we need to solve (\ref{ode1}) for given values of $z$. Let
\[
I=\left(
                 \begin{array}{cc}
                   1 & 0 \\
                   0 & 1 \\
                 \end{array}
               \right), \,\,\,\,
\sigma_3=\left(
                 \begin{array}{rr}
                   1 & 0 \\
                   0 & -1 \\
                 \end{array}
               \right).
\]
Define
\[
N^+(x,z)=m^+(x,z)-I,
\]
and
\[
N^-(x,z)=m^-(x,z)-\sigma_3.
\]
Then (\ref{ode1}) becomes
\begin{align}
N^+_x-[J,N^+]-QN^+=Q,
\label{ode11}
\end{align}
on $[0,\infty)$ with $N^+(\infty)=0$ and
\begin{align}
N^-_x-[J,N^-]-QN^-=Q\sigma_3,
\label{ode12}
\end{align}
on $(-\infty,0]$ with $N^-(-\infty)=0$.
The two equations (\ref{ode11}-\ref{ode12}) are solved column by column using a Chebyshev collocation method \cite{battlestrefethen}. Equation (\ref{ode11}) is solved on $[0,L]$ with a vanishing boundary condition at $x=L$ for sufficiently large $L$ such that the initial values of (\ref{sg}) are smaller than the given tolerance. Similarly,~(\ref{ode12}) is solved on $[-L,0]$ with a vanishing boundary condition at $x=-L$. With the computed solution, the scattering matrix is given by (\ref{scatteringm}):
\[
S(z)=(N^+(0,z)+I)(N^-(0,z)+\sigma_3)^{-1}.
\]
To verify the spectral accuracy of the method, we test it with a known closed-form expression for the reflection coefficient \cite{buckinghammiller}. Consider the initial values
\begin{align}
u(x,0)=2\,\mbox{arccos}(\mbox{tanh}(\epsilon x)),\, u_t(x,0)=2\,\mu\,\mbox{sech}(\epsilon x),
\label{iv}
\end{align}
where $\epsilon$, $\mu$ are real parameters. With proper scaling, these initial values generate solutions to the SG equation in the semiclassical limit as $\epsilon\rightarrow 0$. The reflection coefficient $\rho(z)$ is
\begin{align}
\rho(z)=-\displaystyle\frac{z+(\gamma+\mu)i}{z-(\gamma+\mu)i}\,
\frac{\Gamma\left(\displaystyle\frac{1}{2}+\displaystyle\frac{iE}{\epsilon}\right)
\Gamma\left(1-\displaystyle\frac{\gamma}{2\epsilon}-\displaystyle\frac{iE}{\epsilon}\right)
\Gamma\left(\displaystyle\frac{\gamma}{2\epsilon}-\displaystyle\frac{iE}{\epsilon}\right)}
{\Gamma\left(\displaystyle\frac{1}{2}-\displaystyle\frac{\gamma}{2\epsilon}\right)
\Gamma\left(\displaystyle\frac{1}{2}+\displaystyle\frac{\gamma}{2\epsilon}\right)
\Gamma\left(\displaystyle\frac{1}{2}-\displaystyle\frac{iE}{\epsilon}\right)},
\label{exactrho}
\end{align}
where $E=(z-1/z)/4$, $\gamma=\sqrt{1+\mu^2}$. $\Gamma(z)$ is the gamma function. If $\epsilon=\gamma/(2n+1)$ where $n$ is a non-negative integer, then $\rho(z)\equiv 0$. We choose $\mu=0$, $n=0$, $\epsilon=2$ to have $\gamma/(2n+1)=1 \neq \epsilon$ so that $\rho(z)$ does not vanish. Figure \ref{comprho} shows the numerically computed reflection coefficient on the interval $[-10,10]$ and the spectral decay of the absolute error with respect to the number of collocation points used.
The evaluation of the reflection coefficient off the real line is more difficult since $b(z,t)$ is only guaranteed to have analytic extension in $\mathcal{D}_{\delta}$. Figure \ref{conditionnumber} shows the condition number $K$ when the linear system is solved using a Chebyshev collocation method in the region $0\leq\mbox{\emph{Re}}(z),\mbox{\emph{Im}}(z) \leq 1$. The condition number grows quickly when $z$ gets away from the real axis to the boundary of $\mathcal{D}_{\delta}$. Since the coefficient of the linear term contains $i(z-1/z)/4$ in (\ref{ode11},\ref{ode12}), the condition number $K$ is also related to $\abs{\mbox{Re}(i(z-1/z)/4)}$. On the other hand, with finitely many collocation points near the origin, there exists a narrow band near the origin in which the condition number is moderate. The region with high condition number needs to be avoided and this determines the deformation of the contour near the origin, to be discussed in Section 4.
The computation of the reflection coefficient near the origin can be improved by introducing a new system for $w(x,z)$, called the zero gauge \cite{buckinghammiller,kaup}, with the transformation
\[
w(x,t,z)=i \left(
           \begin{array}{cc}
            \displaystyle\cos\left(u/2\right) & \displaystyle\sin\left(u/2\right) \\
            \displaystyle-\sin\left(u/2\right)& \displaystyle\cos\left(u/2\right)\\
           \end{array}
         \right)m(x,t,z).
\]
The name zero gauge comes from the fact that under this transformation the $1/z$ term in (\ref{Q}) becomes order $z$, and the original system is referred to as infinite gauge. We use the zero gauge to solve for the reflection coefficient for $\abs{z}\leq 1$ and the infinite gauge otherwise. On the other hand, since the reflection coefficient $\rho(z)$ has symmetry (\ref{symmetry1}) and (\ref{symmetry2}), we only need to compute the values of $\rho(z)$ in the upper-half $z-$plane.

\subsection{Computing the zeros of $a(z)$}
\label{sec_poles}
\begin{figure}
  \centering
  \includegraphics[width=0.3\textwidth]{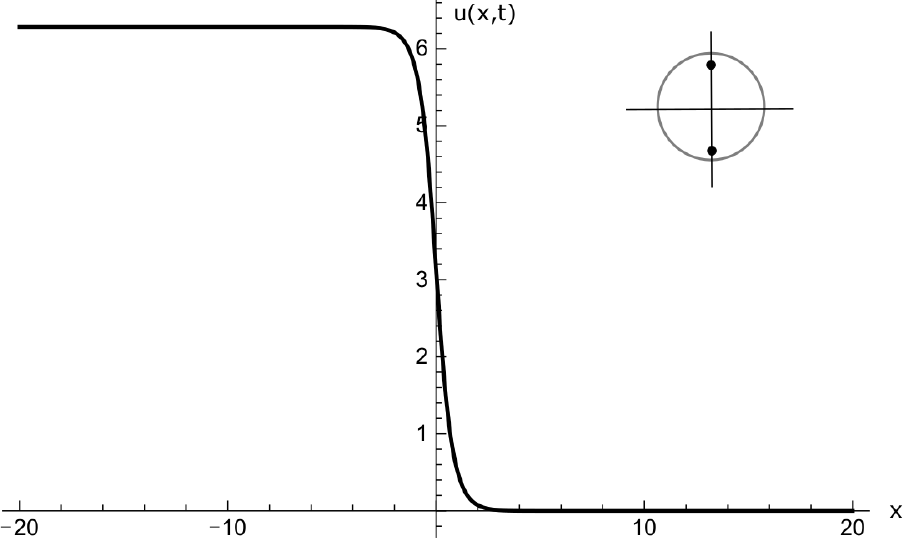}
  \includegraphics[width=0.3\textwidth]{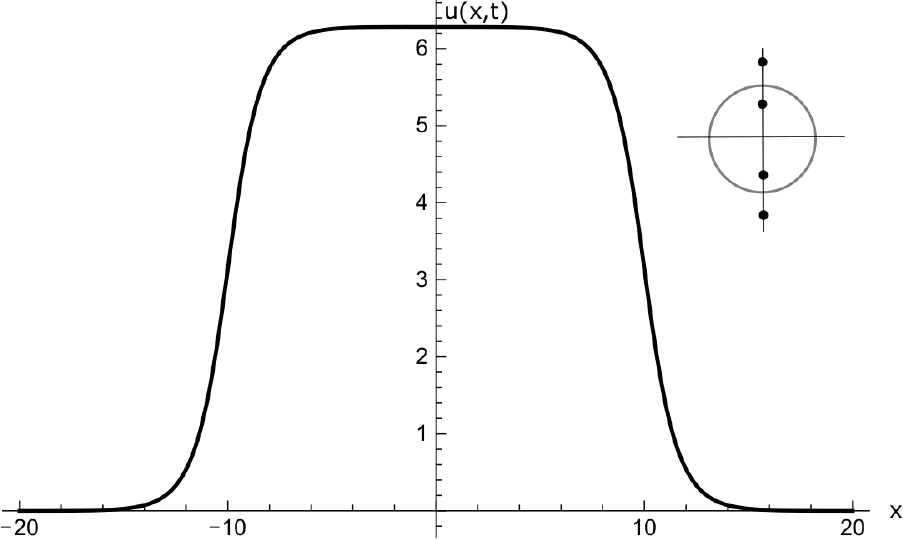}
  \includegraphics[width=0.3\textwidth]{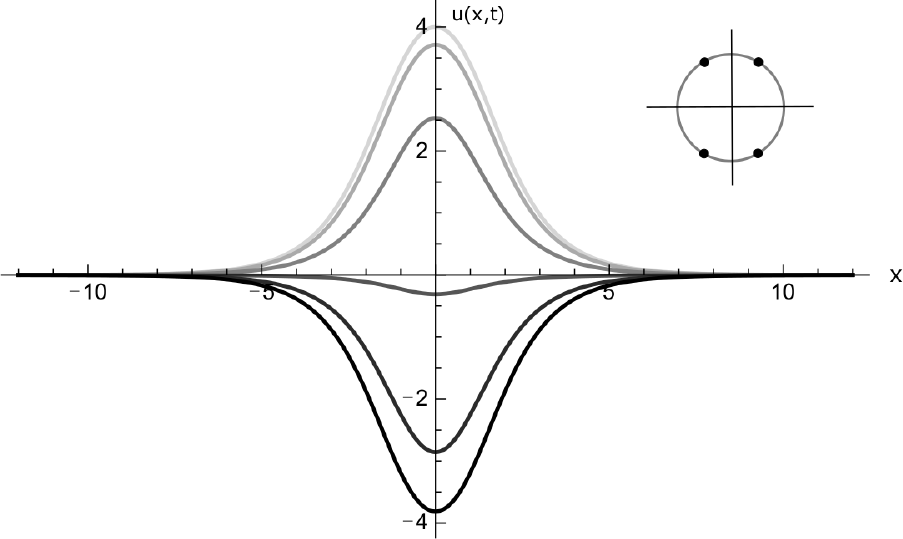}
  \caption{Examples of the three types of soliton solutions $u(x,t)$: (Left) An antikink solution. (Middle) A kink-antikink pair solution. (Right) A sitting breather solution oscillating from upward to downward with profiles in grayscale. The location of the corresponding eigenvalues in the complex $z-$plane is shown in the upper-right corner of each plot. }\label{soliton}
\end{figure}

The zeros of $a(z)$ correspond to the bound states, which are square integrable solutions of (\ref{lpx}). In the KdV equation and the NLS equation, the bound states are obtained by analyzing the discrete spectrum of the Lax operator $X$. Finding the bound states for the SG equation, however, leads to a quadratic eigenvalue problem. It can be written as a standard eigenvalue problem by considering a system of twice the dimension \cite{tisseur},
\begin{align}
 \left(
    \begin{array}{cccc}
      4i\partial_x & i(u_x+u_t) & \cos u & \sin u \\
      i(u_x+u_t) & -4i\partial_x & -\sin u  & \cos u \\
      1 & 0 & 0 & 0 \\
      0 & 1 & 0 & 0 \\
    \end{array}
  \right)
   \left(
   \begin{array}{c}
     \Omega_1 \\
     \Omega_2 \\
   \end{array}
 \right)=z
 \left(
   \begin{array}{c}
     \Omega_1 \\
     \Omega_2 \\
   \end{array}
 \right).
\end{align}
In this case, $\Omega_2=\psi$ and $\Omega_1=z\Omega_2$ are two vectors of size $2 \times 1$. After a change of variable $x\mapsto \tan(s/2)$, Hill's method is applied to compute the eigenvalues of the operator with spectral accuracy \cite{deconinckkut}. We test the accuracy with the initial values in (\ref{iv}). For a general initial values where the eigenvalues are not known, we can check if all the eigenvalues are captured by comparing the reconstructed solution from inverse scattering with the given initial values. Let $n=4$, $\mu=1$, $\gamma=\sqrt{1+\mu^2}=\sqrt{2}$, $\epsilon=\gamma/(2n+1)=\sqrt{2}/9$ in (\ref{iv}). With this form of the initial values, the eigenvalues in the upper half of the complex $z$-plane are classified into three types by the shape of the soliton solutions shown in Figure $\ref{soliton}$:
\begin{enumerate}
\item Antikink: $z=\pm\left(\gamma(\mu)-\mu\right)i$, a single eigenvalue on the positive imaginary axis with its symmetric counterpart on the negative imaginary axis.
\item Kink-antikink pair: $z=\pm i\exp\left(\pm\mbox{arccosh}\left(\gamma(\mu)-2p\epsilon\right)\right)$ with $p\in\mathbb{Z}^+$ such that $1\leq p \leq (\gamma(\mu)-1)/2\epsilon$, a pair of eigenvalues on the positive imaginary axis with one inside and the other outside the unit circle. Two symmetric eigenvalues are on the negative imaginary axis.
\item Breather: $z=\pm\exp(i(\pi/2\pm(\pi/2-\arcsin(\gamma(\mu-2p\epsilon))))$ with $p\in\mathbb{Z}^+$ such that $(\gamma(\mu)-1)/2\epsilon<n\leq \gamma(\mu)/2\epsilon$, four points on the unit circle that are symmetric with respect to both the real and imaginary axis.
\end{enumerate}
\begin{figure}
  \centering
  \includegraphics[width=0.49\textwidth]{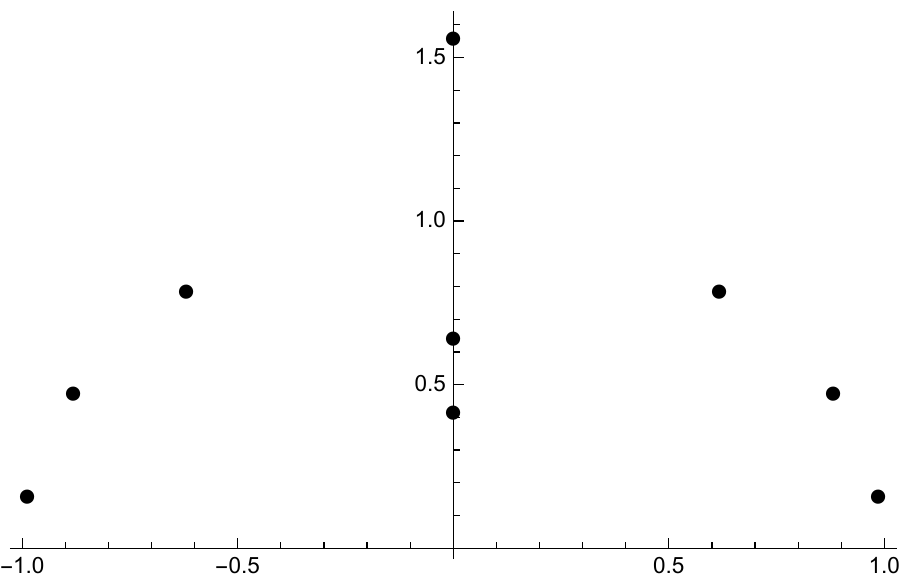}
  \includegraphics[width=0.49\textwidth]{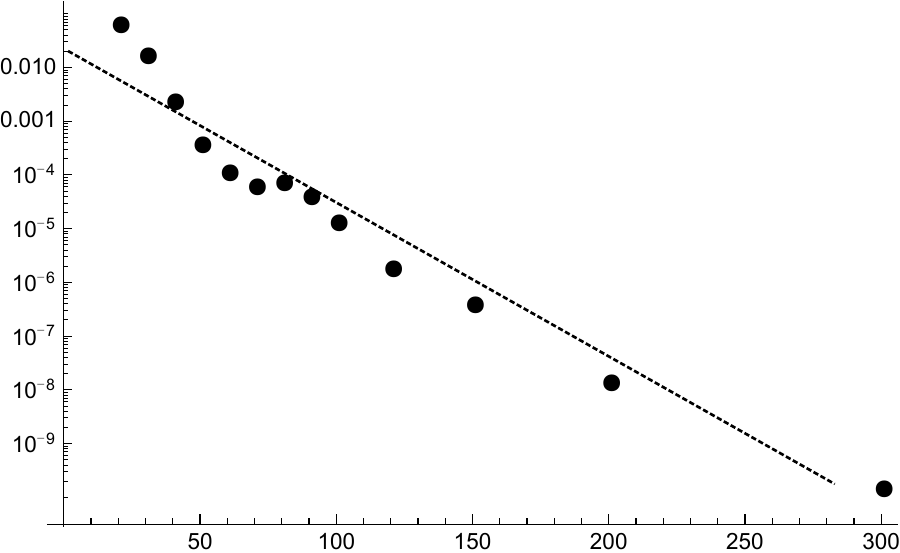}
  \caption{(Left) Location of the eigenvalues in the upper-half $z$-plane. The set consists of $1$ antikink, $1$ kink-antikink pair and $3$ breathers. (Right) The maximum error $y$ versus the number of collocation points $x$. The dashed line a reference straight line.}\label{compeig}
\end{figure}
Figure \ref{compeig} shows the distribution of the eigenvalues in the upper-half $z$-plane and the spectral convergence of the difference between the numerical results using Hill's method and the zeros of the known formula (\ref{exactrho}).
\section{Numerical inverse scattering}
\label{secnist}
\begin{figure}
  \centering
  \includegraphics[width=0.9\textwidth]{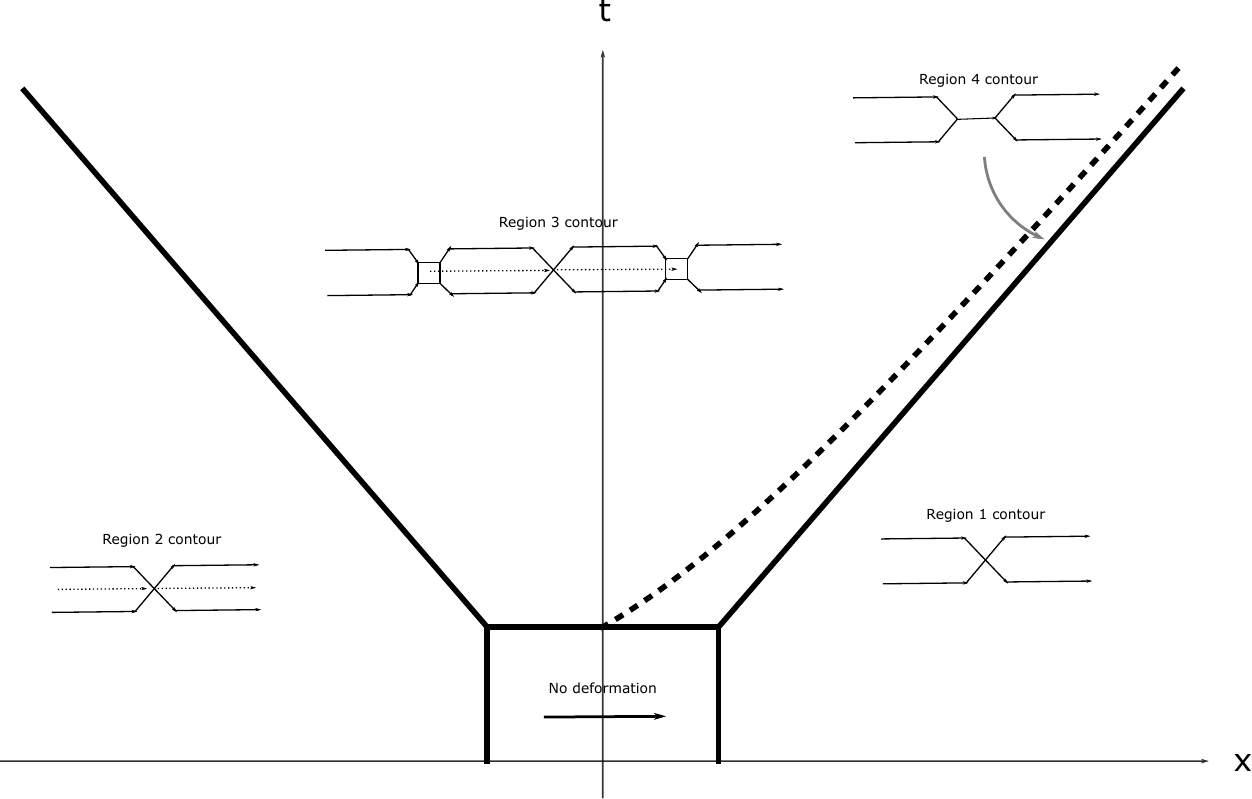}\\
  \caption{Asymptotic regions of the $(x,t)$-plane for the SG equation and the corresponding contour deformation for the IST. Away from the origin, the deformations are restricted in the interior of $\mathcal{D}_{\delta}$. Near the origin, the jump matrix converges to the identity matrix exponentially which allows to use straight lines for the jump contour other than curves tangent to the real axis.} \label{region}
\end{figure}

We solve the RHP (\ref{rhp}) with complex variable $z$ numerically with $x,t$ as parameters. Once the jump matrix can be computed, the packages {\em RHPackage} \cite{rhpackage} by Olver and {\em ISTPackage} \cite{istpackage} by Trogdon are used to solve ~(\ref{rhp}). The idea of the methodology in the packages is that for an RHP denoted by $[G,\Gamma]$, where $G$ is the jump matrix, $\Phi^+=\Phi^-G$, defined on the contour $\Gamma$, we seek a representation of the solution $\Phi$ as
\begin{align}
\Phi=I+\mathcal{C}_{\Gamma} q(z),
\label{sie}
\end{align}
for $q\in L^2(\Gamma)$. Here
\[
\mathcal{C}_{\Gamma} q(z)=\frac{1}{2\pi i} \int_{\Gamma}\frac{ q(s)}{z-s}ds,
\]
is the Cauchy transform of $q(z)$. The Plemelj formula states that $\Phi^+-\Phi^-=q$ \cite{ablowitz3}. Therefore we obtain the singular integral equation (SIE),
\[
q(s)-\mathcal{C}^-_{\Gamma}q(s)(G(s)-I)=G(s)-I,\,\,s\in\Gamma,
\]
where $\mathcal{C}^-_{\Gamma}q(z)$ denotes the non-tangential pointwise limit from the right or the contour $\Gamma$.
We solve the RHP by solving the SIE using the Chebyshev collocation method of Olver \cite{olver}. However, many modifications are required in order to obtain a feasible implementation. As one can see, for general $x$ and $t$, the jump matrix can be highly oscillatory due to the exponential factor in the jump matrix of the RHP (\ref{rhp}), $\theta(z,x,t)=\frac{i}{2}[(z-1/z)x+(z+1/z)t]$. Therefore a large number of collocation points is required to resolve the solution of the RHP. The nonlinear steepest descent method by Deift and Zhou \cite{deiftzhou} provides a solution to this problem by deforming the contour in such a way that the oscillations are turned into exponential decay near the saddle points $\theta'(z_0)=0$. The deformations are different for different regions of $x,t$. For the SG equation, there are four asymptotic regions showing in Figure \ref{region}:
\begin{enumerate}
\item Region 1: Outside the light cone, characterized by $x\geq t$,
\item Region 2: Outside the light cone, characterized by $x\leq -t$,
\item Region 3: Inside the light cone, characterized by $\abs{x}<t$,
\item Region 4: Transition, inside region 3 characterized by $t(t-x)\leq 1$.
\end{enumerate}
After proper deformation of the contours, the jump matrices are exponentially decaying away from the saddle points. Therefore we can truncate the contour if the jump matrix is sufficiently close to the identity matrix, within a given tolerance. In the case of large-parameter asymptotics, the contour becomes localized and the truncation makes the computations more efficient as the parameter increases.
\begin{remark*}
One may expect the deformations in Figure \ref{region} to be symmetric with respect to $x=0$ since the SG equation (\ref{sg}) is invariant under the transformation $x\rightarrow -x$. The asymmetry can be explained in two ways. One reason is that the Lax pair (\ref{lpx}) is not symmetric under $x\rightarrow -x$. The other reason is that the scattering matrix (\ref{scatteringm}) can be interpreted as outputs from sending waves in from $x=-\infty$. If one chooses to define the scattering matrix $\tilde{S}$ by
\begin{align}
\psi^+(x,t,z)\tilde{S}(z,t)=\psi^-(x,t,z),
\label{scatteringmnew}
\end{align}
then the sign in the exponent in the jump matrix changes. As a result, the deformations in Region 1 and Region 2 are swapped and the deformations in Region 3 and Region 4 will be changed owing to the matrix factorization (\ref{mp}) and (\ref{ldu}). We can improve the efficiency of the NIST by eliminating the Region 2 deformation, as discussed in Section \ref{region3}. The machinery of the NIST does not require symmetry in the integrable equation, see the case of the KdV equation \cite{kdv}.
\end{remark*}
\subsection{Region 1: Outside the light cone, characterized by $x\geq t$}
\begin{figure}
  \centering
  \includegraphics[width=\textwidth]{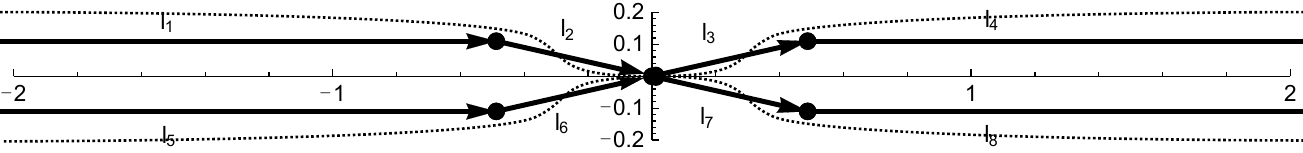}\\
  \caption{The jump contour in the complex plane in the region $x\geq t$ with the $G=MP$ decomposition. The jump contour forms an X shape at the origin. Away from the origin, deformations are inside the dashed lines which are the boundary of $\mathcal{D}_{\delta}$. Near the origin, the straight lines can still be used since the jump matrix is nearly identity.}\label{contour1}
\end{figure}

This is the region where the solution $u(x,t)$ to the SG equation (\ref{sg}) decays to zero faster than any algebraic degree \cite{cheng,huanglenells}. In this region, we introduce the matrix factorization
\begin{align}
G(z,x,t)=M(z,x,t)P(z,x,t),
\label{mp}
\end{align}
where
\[
M(z,x,t)=\left(
    \begin{array}{cc}
      1 & \overline{\rho(\overline{z})}\exp(-\theta(z,x,t)) \\
      0 & 1 \\
    \end{array}
  \right),
\]
\[
P(z,x,t)=\left(
    \begin{array}{cc}
      1 & 0 \\
      \rho(z)\exp(\theta(z,x,t)) & 1 \\
    \end{array}
  \right).
\]
Since
\begin{align}
\mbox{Re}(\theta(z,x,t))=-\frac{\mbox{Im}(z)(x+t)}{2}-\frac{\mbox{Im}(z)(x-t)}{2\abs{z}^2} \left\{
                   \begin{array}{ll}
                     <0, & \hbox{Im $z>0$}, \\
                     >0, & \hbox{Im $z<0$,}
                   \end{array}
                 \right.
\label{sign}
\end{align}
the exponentials in $M$ and $P$ are bounded and decaying if $P$ is defined in the upper-half $z$-plane and $M$ is defined in the lower-half $z$-plane, respectively for $\abs{z}\rightarrow 0,\,\,\infty$ along rays from the origin. Therefore, by a deformation using the lensing \cite{tom} from the real line to the contour in Figure \ref{contour1}, we get a new RHP,
\[
\Phi^+=\left\{
         \begin{array}{ll}
           \Phi^-P, & z\in l_1,l_2,l_3,l_4, \\
           \Phi^-M, & z\in l_5,l_6,l_7,l_8.
         \end{array}
       \right.
\]
The new contour consists of straight-line segments $l_1$-$l_8$. Since $\rho(z)$ is not entire, we are limited in where we can deform to get better decay from the exponential. Away from $z=0$, the width of the strip of analyticity of $\rho(z)$ along the real axis is given by $\delta$ defined in (\ref{schwartz}). The condition number of the collocation method matrix near $z=0$ for (\ref{ode1}) is shown in Figure \ref{conditionnumber}, the level sets are used to determine the deformation. We first pick $\nu<\delta/2$ to determine the $y-$coordinate of the horizontal segments $l_1,l_4,l_5,l_8$. Then $\abs{\mbox{Re}(i(z-1/z)/4)}=2\nu$ determines the circle centered at $z=i/8\nu$ with radius $1/8\nu$ from which we can solve for the intersection of the circle with the straight horizontal line $y=\nu$. For convenience, the arc from $z=0$ to $y=\nu$ is replaced by a straight line in our experiments since $\nu<0.4$ is small. The deformations near $z=0$ in other asymptotic regions with the exception of the transition region are determined using the same method.

\subsection{Region 2: Outside the light cone, characterized by $x\leq -t$}
\begin{figure}
  \centering
  \includegraphics[width=\textwidth]{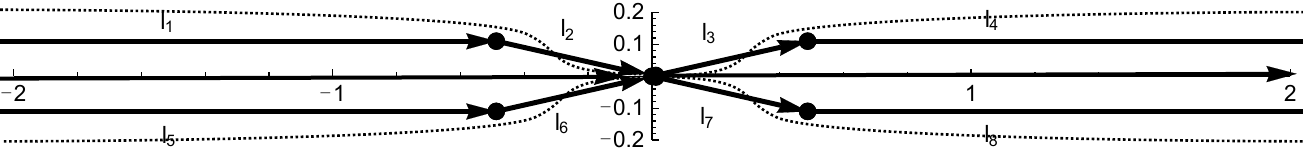}\\
  \caption{The jump contour in the complex plane in the region $x\leq -t$ with the $G=LDU$ decomposition. Away from the origin, deformations are inside the dashed lines which are the boundary of $\mathcal{D}_{\delta}$. Near the origin, the straight lines can still be used since the jump matrix is nearly identity. The jump on the real line introduces large oscillations in the solution and needs to be removed to obtain uniform accuracy. After the removal of the jump on the real line, the contour is of the same shape as in Figure \ref{contour1}. }
  \label{contour2a}
\end{figure}

This is the other region where $u(x,t)$ decays to zero faster than any algebraic degree \cite{cheng,huanglenells}. In this region, the sign of the real part of $\theta$ is the opposite of (\ref{sign}). A different matrix factorization is needed:
\begin{align}
G=LDU,
\label{ldu}
\end{align}
where
\[
L(z,x,t)=\left(
    \begin{array}{cc}
      1 & 0 \\
      \frac{\rho(z)}{\tau(z)}\exp(\theta(z,x,t)) & 1 \\
    \end{array}
  \right),
\]
\[
U(z,x,t)=\left(
    \begin{array}{cc}
      1 & \frac{\overline{\rho(\overline{z})}}{\tau(z)}\exp(-\theta(z,x,t)) \\
      0 & 1 \\
    \end{array}
  \right),
\]
\[
D(z)=\left(
       \begin{array}{cc}
         \tau(z) & 0 \\
         0 & \frac{1}{\tau(z)} \\
       \end{array}
     \right),
\]
and $\tau(z)=1+\overline{\rho(\overline{z})}\rho(z)$. The jump function $L$ contains a decaying exponential if it is defined in the lower-half $z$-plane while the jump function $U$ has decaying exponential if it is defined in the upper-half plane, respectively for $\abs{z}\rightarrow 0,\,\,\infty$ along rays from the origin. We get an RHP as shown in Figure \ref{contour2a}:
\[
\Phi^+=\left\{
         \begin{array}{ll}
           \Phi^-U, & z\in l_1,l_2,l_3,l_4, \\
           \Phi^-D, & z\in \mathbb{R},\\
            \Phi^-L, & z\in l_5,l_6,l_7,l_8.
         \end{array}
       \right.
\]
Similar to the discussion in region 1, straight lines are used since the jump matrix converges to the identity matrix exponentially.
To obtain uniform accuracy, all the jump matrices need to approach the identity matrix away from the saddle points \cite{tom}. For the diagonal jump matrix $D$, we can write down the exact solution $\Delta(z)$ to the RHP,
\[
\Delta^+=\Delta^-D, \,\, z\in \mathbb{R},\,\, \lim_{z\rightarrow \infty} \Delta(z)=I,
\]
where
\[
\Delta(z)=\left(
             \begin{array}{cc}
               \delta(z) & 0 \\
               0 & 1/\delta(z) \\
             \end{array}
           \right),
\]
and
\[
\delta(z)=\exp\left(\frac{1}{2\pi i}\int_{-\infty}^{\infty} \frac{\log (\tau(s))}{s-z}ds\right).
\]
Hence, using the mapping $\Phi^+ \mapsto \Phi^+ \Delta^{-1}$ and $\Phi^- \mapsto \Phi^- \Delta^{-1}$ we can remove the jump $D$ on $\mathbb{R}$ and the contour becomes the same as in Figure \ref{contour1} with
\[
\Phi^+=\left\{
         \begin{array}{ll}
           \Phi^-\Delta U \Delta^{-1}, & z\in l_1,l_2,l_3,l_4, \\
           \Phi^-\Delta L \Delta^{-1}, & z\in l_5,l_6,l_7,l_8.
         \end{array}
       \right.
\]

\subsection{Region 3: Inside the light cone, characterized by $\abs{x}<t$}
\label{region3}
\begin{figure}
  \centering
  \includegraphics[width=\textwidth]{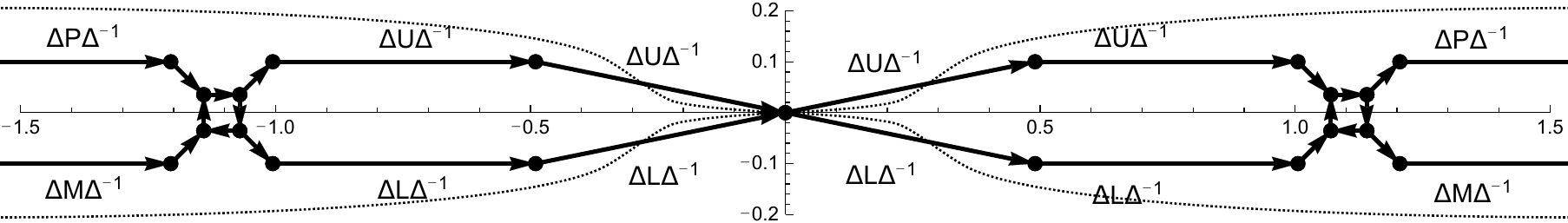}\\
  \caption{The jump contour in the complex plane in the region $\abs{x}<t$ with the jump functions labeled to each segment. Away from the origin deformations are inside the dashed lines which are the boundary of $\mathcal{D}_{\delta}$. Near the origin, the straight lines can still be used since the jump matrix is nearly identity.}\label{contour3}
\end{figure}

\begin{figure}
  \centering
  \includegraphics[width=0.45\textwidth]{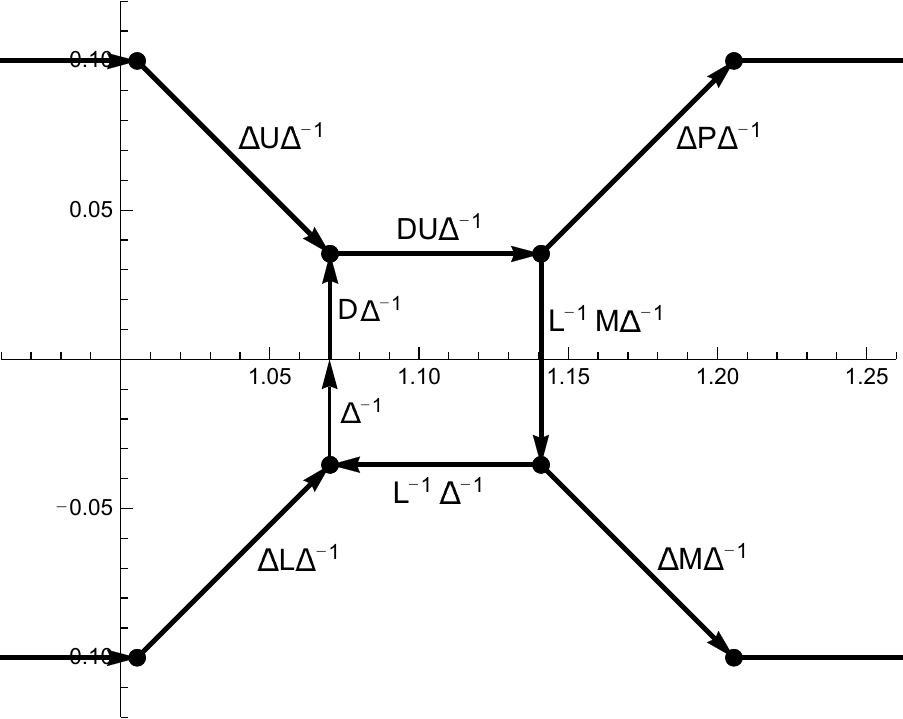}\\
  \caption{A zoom-in of the contour in Figure \ref{contour3} at the right saddle point. The four oriented contours pointed at/from the square are along the direction of steepest descent ($\pi/4$ from real axis). The center of the square is the saddle point $z_0$.) }\label{contour4}
\end{figure}

The deformation in Region 3 is more complicated and contains types of deformations discussed in the previous two regions. For convenience, we denote the jump matrices on the contours in Figures \ref{contour3}, \ref{contour4} and \ref{contour5}. For instance, a jump matrix $G$ next to the oriented contour means that the solution to the RHP $\Phi$ satisfies $\Phi^+=\Phi^- G$. When $(x,t)$ is inside the light cone, we have two real saddle points at $\pm z_0=\pm \sqrt{(t-x)/(t+x)}$ satisfying $\theta'(z_0,x,t)=0$. The two saddle points are moving away from the origin and are unbounded when $x\rightarrow -t$. They approach the origin when $x\rightarrow t$. Note that in Region 1 and Region 2, $\abs{x}\geq t$, the two saddle points are purely imaginary. Near the two real saddle points,
 \begin{align}
 \theta(z,x,t)=\frac{i (t-x)}{\sqrt{\frac{t-x}{t+x}}}+\frac{i (t+x) \left(z-\sqrt{\frac{t-x}{t+x}}\right)^2}{2 \sqrt{\frac{t-x}{t+x}}}+O\left(\left(z-\sqrt{\frac{t-x}{t+x}}\right)^3\right).
 \label{thetaexp}
 \end{align}
 To get decay from the quadratic term, for $\mbox{Re}(z)>z_0$ and $\mbox{Re}(z)<-z_0$, we need the $G=MP$ factorization and for $-z_0<\mbox{Re}(z)<z_0$ we need the $G=LDU$ factorization. Figure \ref{contour3} shows the deformation with saddle points at $\pm z_0\approx \pm 1.1$. Furthermore, to get uniform accuracy, we need to remove the jump $D(z)$ on $(-z_0,z_0)$ by introducing the RHP,
\[
\Delta^+=\Delta^-D, \,\, z\in (-z_0,z_0),\,\, \lim_{z\rightarrow \infty} \Delta(z)=I,
\]
with exact solution
\[
\Delta(z;z_0)=\left(
             \begin{array}{cc}
               \delta(z;z_0) & 0 \\
               0 & 1/\delta(z;z_0), \\
             \end{array}
           \right),
\]
and
\[
\delta(z;z_0)=\exp\left(\frac{1}{2\pi i}\int_{-z_0}^{z_0} \frac{\log (\tau(s))}{s-z}ds\right).
\]
In this case, $\delta(z;z_0)$ has singularities at $\pm z_0$. To avoid using contours passing through the singularity, we introduce a new square contour centered at the singularity as in Figure \ref{contour4}. The length of the side is on the order of $\sqrt{(t+x)^3}/\sqrt{t-x}$ determined by the coefficient of the  quadratic term in (\ref{thetaexp}). When $x\rightarrow -t$, since $\rho(z_0)$ decays to zero quickly, the contour near the saddle points may be truncated and therefore the contour degenerates to the contour in region 2. From the expansion of $\theta(z,x,t)$ (\ref{thetaexp}), we can see that the localization depends on the absolute value of the coefficient of the quadratic term. When $t+x\approx 0$, the jump matrix may still be very oscillatory due to the insufficient decay in $z$. However, since the constant term in (\ref{thetaexp}) determines the overall amplitude of the jump matrix which will be truncated if it gets too small, the oscillation cannot become arbitrarily large. Two techniques are used to deal with the intermediate oscillatory case. A technique to compute the solution for $x<0$ is to use a reflected initial values $v(x,0)=u(-x,0)$ which gives a new set of scattering data but only requires the deformation of the contour in the $x>0$ case for $v(x,t)=u(-x,t)$. The other technique is to precompute the reflection coefficient along the contour on a coarser grid and use the interpolants to construct the jump matrix. This is effective because the oscillation is mostly introduced by $\theta(z,x,t)$ but the refection coefficient itself is smooth as in Figure \ref{comprho}. On the other hand, when $x\rightarrow t$, the collision of the two saddle points $\pm z_0$ at the origin results in a contour in region 1, and the $G=LDU$ factorization is indeed not necessary in the transition region, as we now demonstrate.
\subsection{Region 4: Transition, inside region 3 characterized by $t(t-x)\leq 1$}
\begin{figure}
  \centering
  \includegraphics[width=\textwidth]{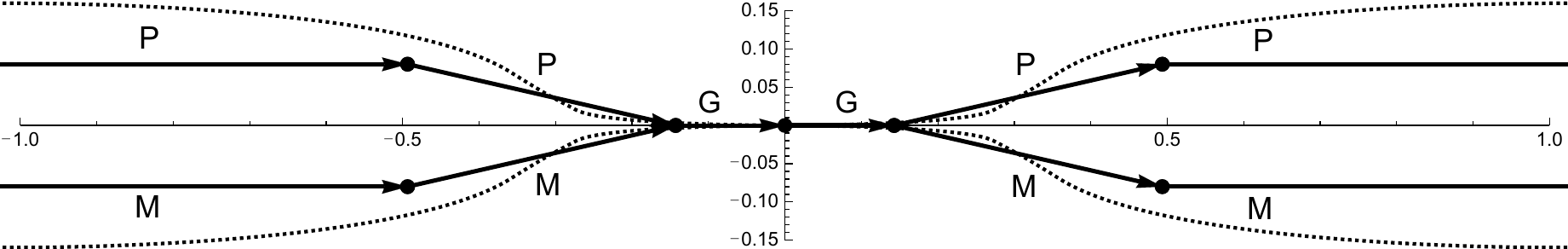}\\
  \caption{The jump contour in the complex plane in the transition region. When the saddle points are near the origin, the $G=LDU$ decomposition is collapsed to the real line. Away from the origin, deformations are inside the dashed lines which are the boundary of $\mathcal{D}_{\delta}$. Near the origin, the straight lines can still be used since the jump matrix is nearly identity.}\label{contour5}
\end{figure}

When $t-x=\varepsilon>0$ is small, the saddle points approach the origin. This is classified as a transition region in \cite{huanglenells}. Consider $z_0<1/t$ and  $-z_0<z<z_0$, then
\[
\abs{\theta(z,x,t)}=\abs{zt}+\frac{1}{2}\abs{(z-1/z)(t-x)}\leq 1+\frac{1}{2}\abs{(z-1/z)}\epsilon.
\]
The oscillation is indeed controlled between the two saddle points when their distance is sufficiently close. In this case, from Theorem 2.74 and Corollary 2.75 in \cite{tom}, the Sobolev norm of the solution $q$ to the SIE (\ref{sie}) can be bounded uniformly in $x$ and $t$ since $\theta(z,x,t)$ is bounded independent of $x$ and $t$. Therefore there is no need to use the $G=LDU$ factorization and we can collapse the contour back to the real line as in Figure \ref{contour5}.

\section{Numerical results}
We present some numerical examples and tests of the NIST.
\subsection{Propagation of dispersive waves}
\begin{figure}
  \centering
  \includegraphics[width=\textwidth]{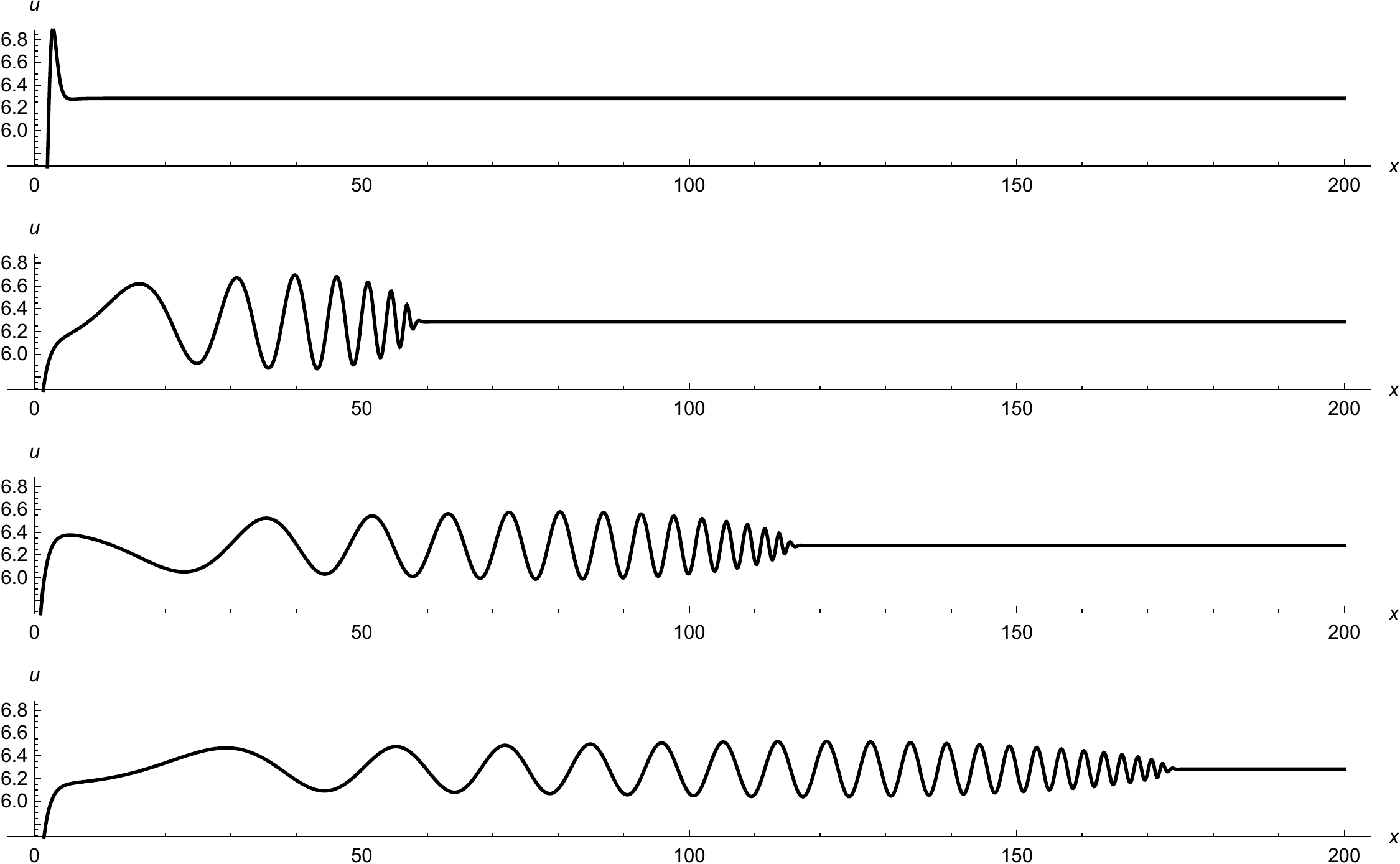}
  \caption{The propagation of dispersive waves to the right from a perturbed stationary kink located at $x=0$ when $t=2.5,\,60,\,120,\,180$. }\label{sol2}
\end{figure}

Let
\[
u_s(x,t)=4 \mbox{arctan}\left(e^x\right),
\]
be a one-soliton stationary kink solution. We choose the initial values to be a (not small) perturbation of $u_s(x,0)$,
\[
u(x,0)=u_s(x,0)+5\,\mbox{sech}^2(x),
\]
\[
u_t(x,0)=0.
\]
The magnitude of the perturbation is chosen to introduce large dispersion while the number of eigenvalues in the scattering problem does not change. Figure \ref{sol2} shows dispersive waves generated by the perturbation from the kink at $t=2.5,\,60,\,120,\,180$. Due to the oscillations and the different scales, the solution is difficult to obtain with traditional numerical methods while maintaining high accuracy for long time.

\subsection{Recovery of the initial values}
\label{sec_recover}
\begin{figure}
  \centering
  \includegraphics[width=0.49\textwidth]{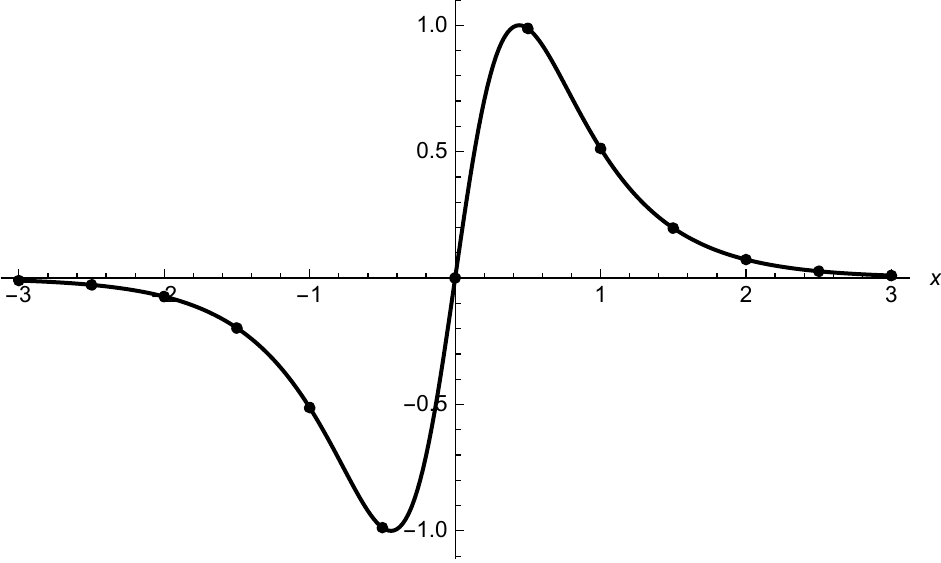}
  \includegraphics[width=0.49\textwidth]{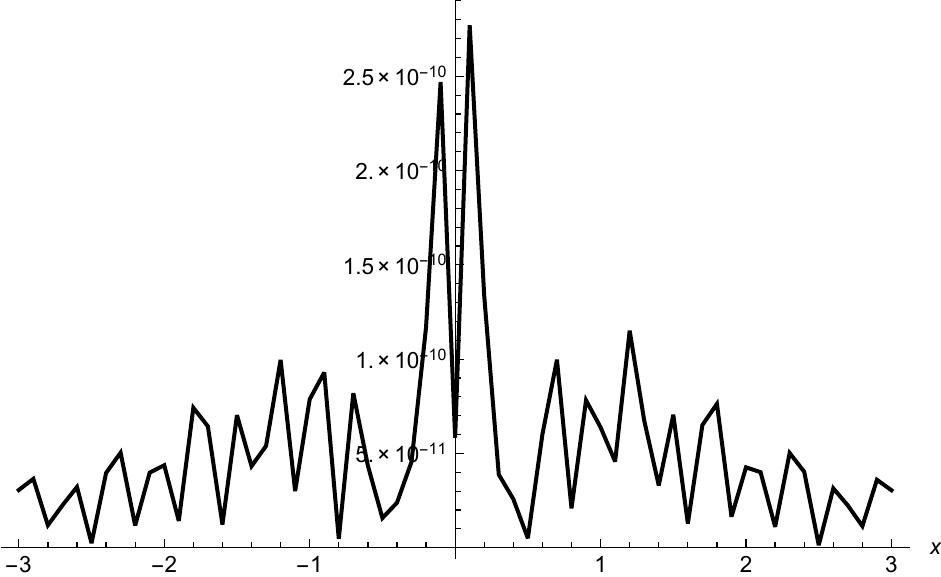}\\
  \caption{(Left) Solid: The exact initial values $\sin(u(x,0))$, Dots: The computed solution. (Right) The absolute error. }\label{sol1}
\end{figure}

\begin{figure}
  \centering
  \includegraphics[width=0.5\textwidth]{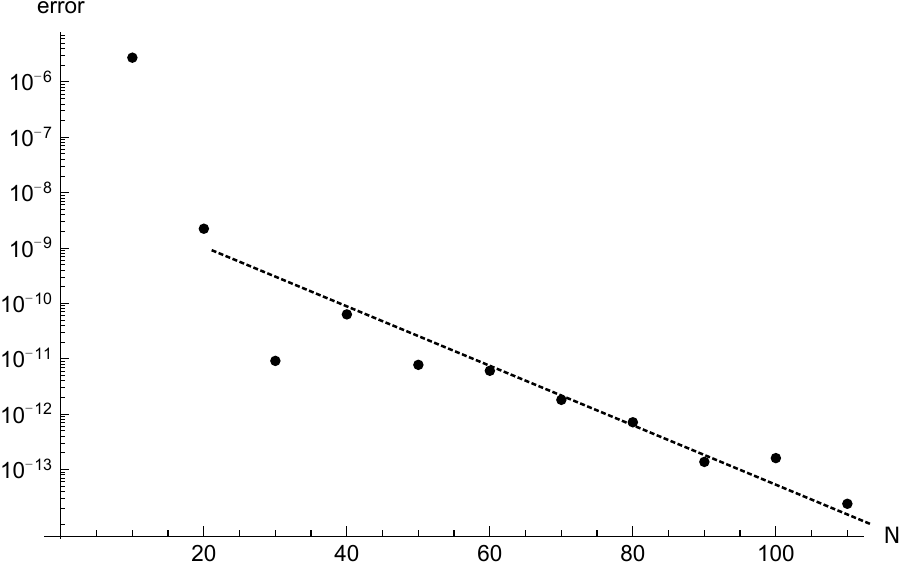}\\
  \caption{Log-linear plot of the error {\em vs}. number of collocation points for $u(1.5,1)$. The dashed line is a reference straight line.}\label{conv}
\end{figure}

As mentioned in Section \ref{sec_poles}, we can solve the RHP at $t=0$ and compare it with the known initial values to verify that all eigenvalues have been computed. This is one way to check the accuracy when the exact solution is not known. Figure \ref{sol1} shows the computed initial values using the NIST on the left at $t=0$ with initial values (\ref{iv}) and $\mu=0$, $n=0$, $\epsilon=2$. The absolute error is shown on the right and is on the order of $10^{-10}$. Although it is a triviality for numerical methods like finite differences to check the initial values, in the NIST, computing $u(x,0)$ requires going through the entire procedure of direct scattering and inverse scattering. Computing short time solutions in fact costs more compared with computing the solutions for long time since the contour becomes more localized when $t$ is large \cite{olvertrogdon}.
With the same initial value, the spectral convergence at $x=1.5$, $t=1$ is verified in Figure \ref{conv}. The error is measured by the difference of two numerical solutions with a different number of collocation points, $e_{N_i}=\abs{u_{N_{i+1}}-u_{N_{i}}}$. Linear behavior in the log plot indicates spectral convergence.

\subsection{Evolution of multi-soliton solutions with perturbations}
\begin{figure}
  \centering
  \includegraphics[width=0.49\textwidth]{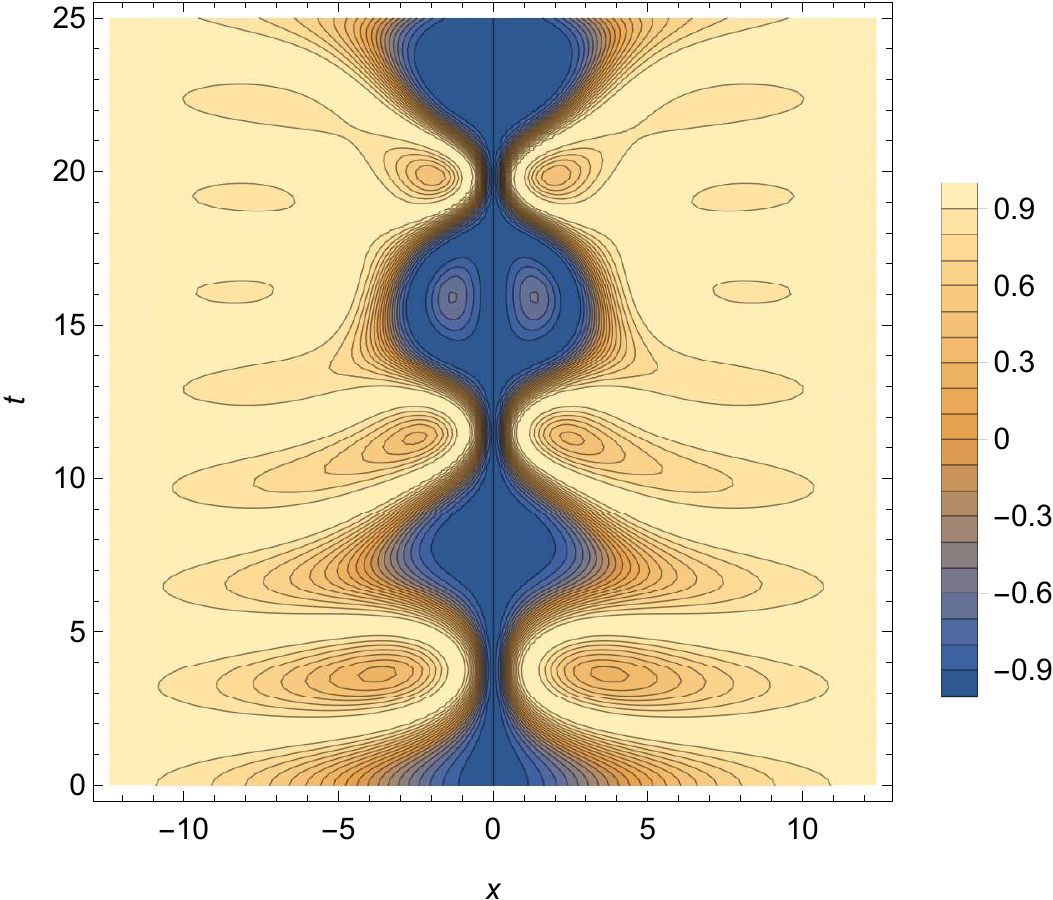}
  \includegraphics[width=0.49\textwidth]{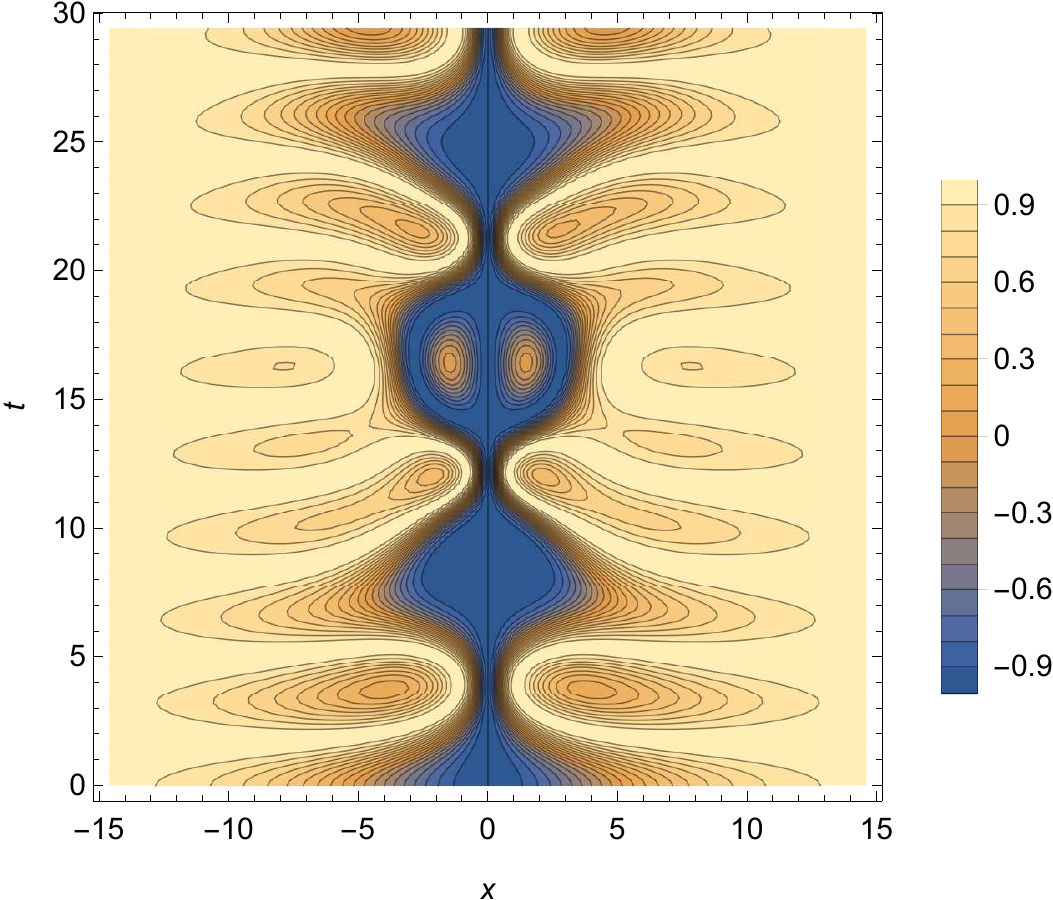}\\
  \caption{(Left) The numerical solution $\cos(u)$ with initial values (\ref{iv}), $\mu=0$, $n=2$ and $\epsilon=0.2$. In this case the reflection coefficient vanishes. $\rho(z)\equiv 0$. The solution consists of several solitons as discussed in Section \ref{sec_poles}. (Right) The numerical solution with initial values (\ref{iv}), $\mu=0$, $n=2$ and $\epsilon=0.17$. In this case, the solitons are perturbed and small dispersive waves exists since $\rho(z)\not \equiv 0$ but dispersive waves are hardly seen due to the small amplitude. The domain is the same in the two plots with scaling $\epsilon x\in [-2.5,2.5]$, $\epsilon t\in [0,5]$.}\label{contoursc}
\end{figure}

\begin{figure}
  \centering
  \includegraphics[width=0.39\textwidth]{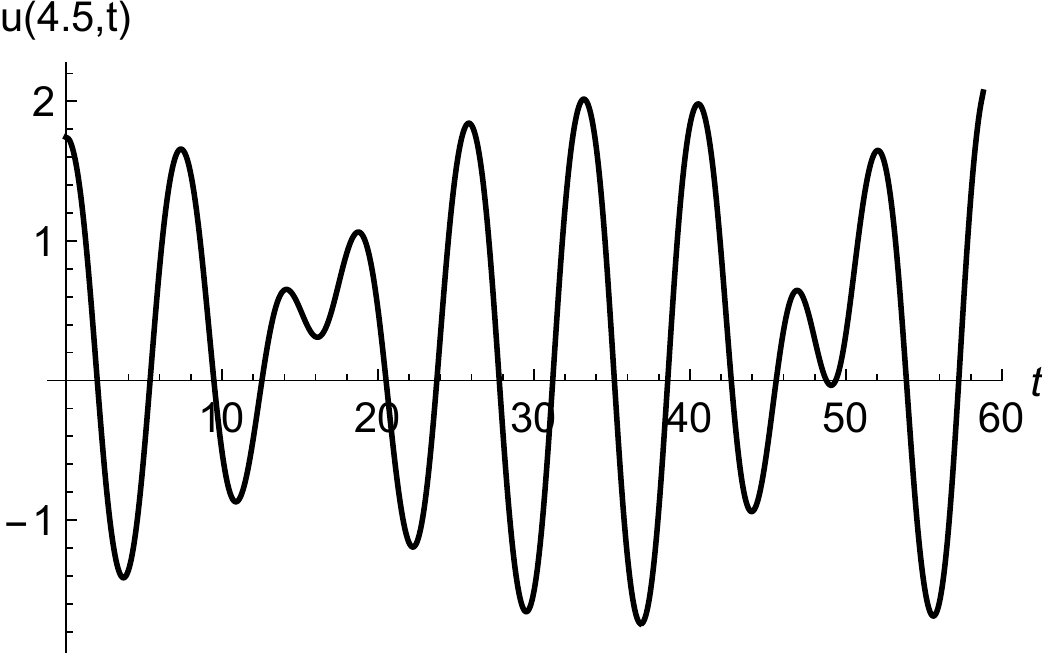}
  \hspace{1cm}
  \includegraphics[width=0.52\textwidth]{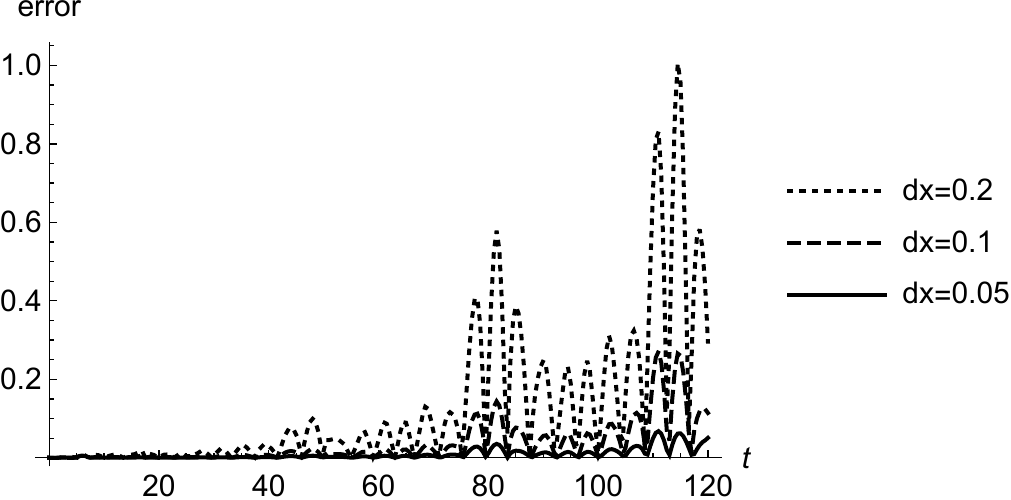}\\
  \caption{(Left) The numerical solution by the NIST $u(4.5,t)$. (Right) The absolute error compared with a second-order finite-difference method with various grid sizes at $x=4.5$, $t\in [0,120]$.}\label{compfd}
\end{figure}

In \cite{buckinghammiller}, the initial values (\ref{iv}) with $\epsilon=\gamma/(2n+1)$ are used so that one can compute pure soliton solutions by solving an algebraic system in the semiclassical limit $\epsilon\rightarrow 0$. If the condition $\epsilon=\gamma/(2n+1)$ is not satisfied, an RHP has to be solved due to the non-zero reflection coefficient, and the algebraic method does not apply. One expects that the non-zero reflection coefficient perturbs the solution only by a small amount, and the pure soliton solution approximates the general solution as $\epsilon\rightarrow 0$. In Figure \ref{contoursc}, we show the pure soliton solution with $\mu=0, n=2, \epsilon=0.2$ in the left panel. The perturbed solution with $\mu=0, n=2, \epsilon=0.17$ is in the right panel using the NIST. In both cases the reflection coefficient has five poles on the unit circle in the upper-half plane. In the study of the semiclassical limit $\epsilon \rightarrow 0$, the scaling $X=\epsilon x$ and $T=\epsilon t$ is relevant. This transforms the SG equation to
\[
  \epsilon^2 U_{TT}-\epsilon^2 U_{XX} + \sin(U)=0.
\]
Therefore the domain $\epsilon x\in [-2.5,2.5]$, $\epsilon t\in [0,5]$ remains the same in Figure \ref{contoursc}. The two plots are similar with a small difference of contour lines near the center of the plots. 
We remark that computing a 2D contour plot is not efficient using the NIST since the advantage of the NIST is that the solution is computed at specified $(x,t)$ without time-stepping. All points in the evolution are required in a 2D plot. However, it is competitive to use the NIST in this case if one wants accurate solutions containing dispersive waves for large time. In Figure \ref{compfd}, we compare the numerical solution by the NIST with a standard centered-difference method along the line $x=4.5$. To prevent introducing error due to domain truncation, we choose the domain from $[-7.5/\epsilon,7.5/\epsilon]$. The left plot shows the oscillatory evolution of the solution mostly due to the existence of breathers with the error on the order of $10^{-8}$. The dispersive waves are not large enough to be observed in the plot. In the right panel of Figure \ref{compfd}, three grid sizes $\Delta x=0.05, 0.1, 0.2$ are used with the time step $\Delta t=\Delta x/2$ to satisfy the stability condition. These step sizes are sufficient to resolve the oscillations in the plot but the error is on the order of $0.01$ for small time and seems to grow to order $1$ linearly. As a result, an extremely dense grid is required to get the solution with error smaller than $10^{-8}$.
\subsection{Uniform convergence for large t and x}
\begin{figure}
  \centering
  \includegraphics[width=0.6\textwidth]{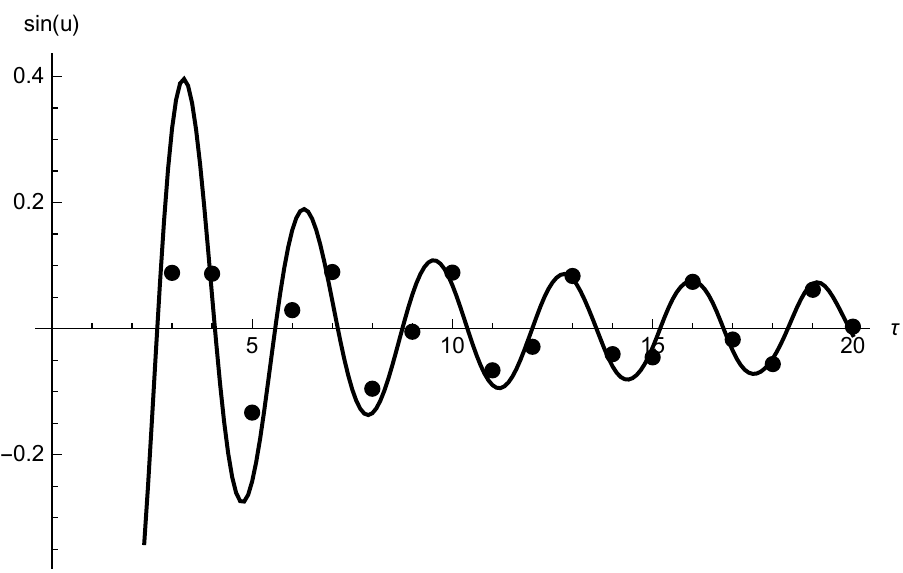}\\
  \caption{Comparison of the numerical solution $\sin(u(\tau))$ (Solid) to the asymptotic formula given by (\ref{asympsex}) (Dots) for $\tau\in[0,20]$, $z_0=\sqrt{1/2}$. }\label{asymp1}
\end{figure}

\begin{figure}
  \centering
  \includegraphics[width=0.6\textwidth]{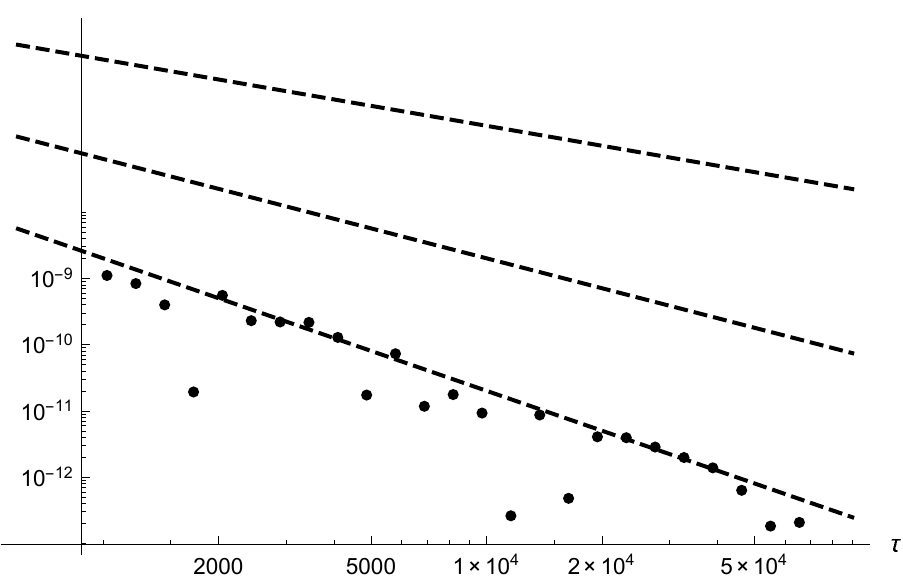}\\
  \caption{The absolute error between the numerical solution $\cos(u(\tau))$ and the asymptotic formula.  (Dot) The error/log($\tau$). (Dashed) Auxiliary lines with slope $-1,-1.5,-2$. The least square fit of the error/log($\tau$) has slope $-1.95$, as opposed to $-1.5$, predicted by \cite{cheng}.}\label{asymp2c}
\end{figure}

\begin{figure}
  \centering
  \includegraphics[width=0.6\textwidth]{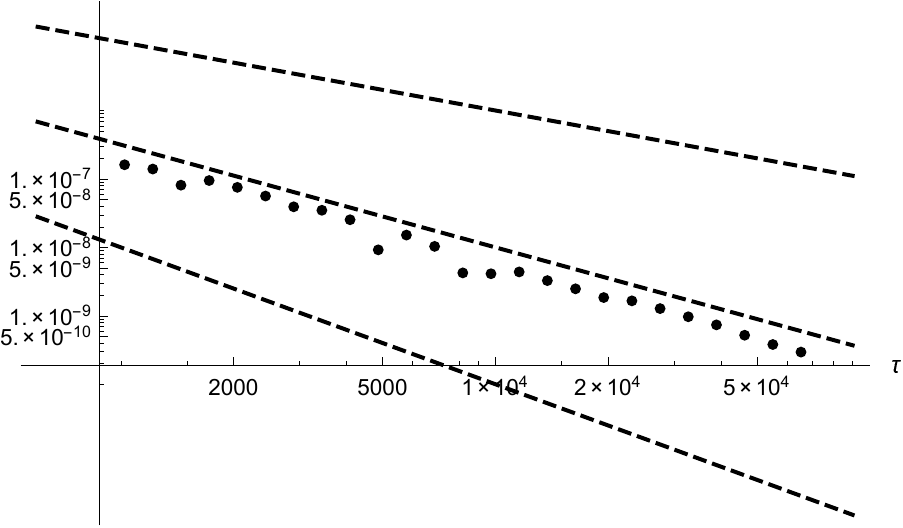}\\
  \caption{The absolute error between the numerical solution $\sin(u(\tau))$ and the asymptotic formula.  (Dot) The error/log($\tau$). (Dashed) Auxiliary lines with slope $-1,-1.5,-2$. The least square fit of the error/log($\tau$) has slope $-1.53$, as opposed to $-1$, predicted by \cite{cheng}.}\label{asymp2s}
\end{figure}

Cheng, Venakides and Zhou \cite{cheng} studied the long-time asymptotics for solitonless initial values. Outside the light cone, the solution decays to zero spectrally. Inside the light cone, let
\[
z_0=\sqrt{\frac{t-x}{t+x}} ,\,\,\,
\tau=\frac{tz_0}{1+z_0^2},
\]
then as $\tau\rightarrow\infty$,
\begin{align}
\cos(u)-1=-\frac{4\abs{\nu(z_0)}}{\tau} \cos^2(2\tau + \nu(z_0)\log(8\tau)+\beta(z_0))+O\left(C(z_0)\frac{\log(\tau)}{\tau^{3/2}}\right),\\
\sin(u)=\sqrt{\frac{8\abs{\nu(z_0)}}{\tau}} \cos(2\tau + \nu(z_0)\log(8\tau)+\beta(z_0))+O\left(C(z_0)\frac{\log(\tau)}{\tau}\right),
\label{asympsex}
\end{align}
where
\[
\nu(z_0)=-\frac{1}{2\pi}\log(1+\abs{\rho(z_0)}^2),
\]
\[
\beta(z_0)=-\arg(\Gamma(\nu(z_0)i))-\arg(\overline{\rho(z_0)})+\frac{\pi}{4}-\frac{1}{\pi}\int^{z_0}_{-z_0}
\log(z_0-s)d\log(1+\abs{\rho(s)}^2).
\]
Here, $\Gamma(\cdot)$ denotes the Gamma function and $C(z_0)$ decays faster than any power of $z_0$ as $z_0\rightarrow \infty$.
In Figures \ref{asymp1}, \ref{asymp2c}, \ref{asymp2s}, we compare the numerical solution with asymptotic formula inside the light cone region $z_0=\sqrt{1/2}$ with initial values $u(x,0)=\mbox{sech}^2(x)$ and $u_t(x,0)=0$. The observed orders of the correction terms are one half order smaller than the order given by the asymptotic formula. This is true for all initial values we have tested, which indicates a possible refinement of the estimates of the correction terms in \cite{cheng}.

\subsection{Comparison with the auto-B\"{a}cklund transformation}
\begin{figure}
  \centering
  \includegraphics[width=0.45\textwidth]{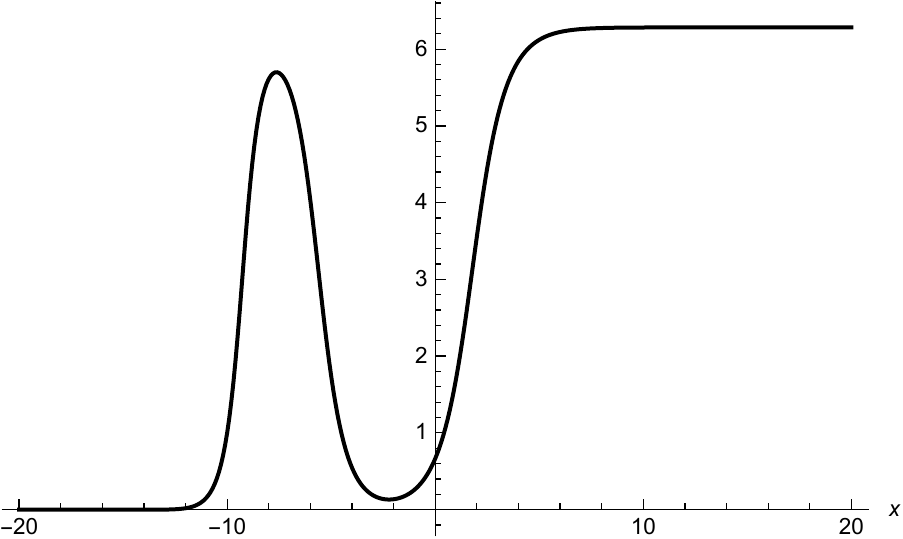}
  \includegraphics[width=0.45\textwidth]{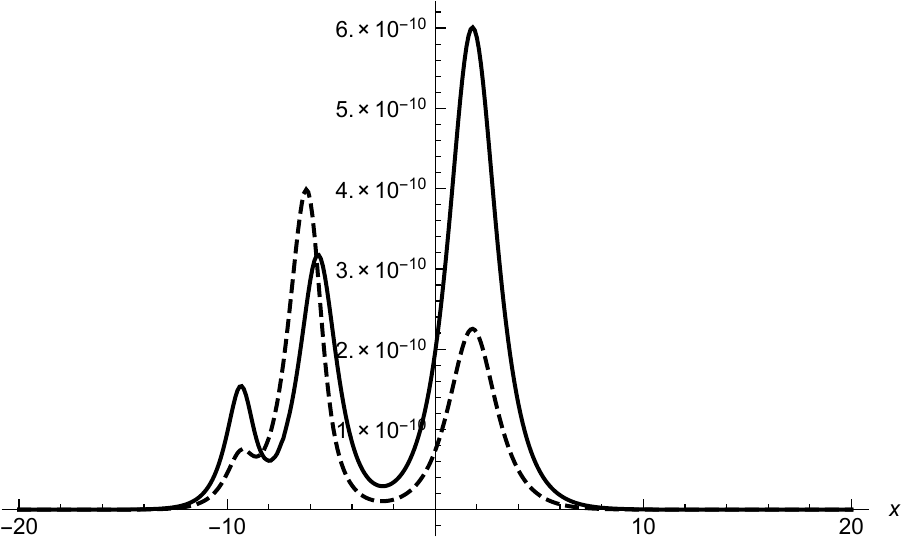}\\
  \caption{Left: The exact three-soliton solution at $t=10$. Right: The error compared with the exact solution (solid) and the residual of (\ref{bt}) (dashed).}\label{bt1}
\end{figure}

An important property of integrable systems is that they have a B\"{a}cklund transformation. Two solutions $u(x,t)$ and $v(x,t)$ to (\ref{sg}) satisfy the auto-B\"{a}cklund transformation if they satisfy the following equations \cite{doddbullough},
\alpheqn
\begin{align}
u_x+u_t=&v_x+v_t+2k\sin\left(\frac{u+v}{2}\right),
\label{abt1}
\\
u_x-u_t=&-(v_x-v_t)+\frac{2}{k}\sin\left(\frac{u-v}{2}\right).
\label{abt2}
\end{align}
\resetalpheqn
By choosing the parameter $k$, we can find a new solution $u$ from a known solution $v$ using the transformation. The effect of the transformation is the addition or removal of one zero of $a(z)$ in the upper-half plane. By cancellation of the derivative terms in the transformation (\ref{abt1}, \ref{abt2}) using parameters $k_1,k_2$ in different order, we get an algebraic consistency condition among four solutions $r,u,v,w$ to (\ref{sg}),
\begin{align}
\tan\left(\frac{w-r}{4}\right)=\frac{k_2+k_1}{k_2-k_1}\tan\left(\frac{u-v}{4}\right),
\label{bt}
\end{align}
where $v$ is obtained using (\ref{bt}) with $r$, $k_1$ and $u$ is obtained using (\ref{abt1}, \ref{abt2}) with $r$, $k_2$.
Since the trivial solution $v=0$ satisfies (\ref{sg}), we obtain three one-soliton solutions $u_j$ from (\ref{abt1}, \ref{abt2}),
\begin{align}
u_j=4\arctan \left(\exp\left(k_j x+\frac{1}{k_j}t\right)\right),\, k_j=j,\, j=1,2,3.
\label{1soliton}
\end{align}
and construct a three-soliton solution by using (\ref{bt}) repeatedly. Figure \ref{bt1} shows the three-soliton solution at $t=10$ on the left. The error comparing with the exact solution is shown in a solid line in the right plot. The residual of (\ref{bt}) is examined by computing all the soliton solutions from their initial values to $t=10$ independently. Both the absolute error and the residual stay small uniformly in both $t$ and $x$ in the computation.


\section*{Acknowledgements}
The authors gratefully acknowledge support from the National Science Foundation under grant NSF-DMS-1522677(BD,XY). Any opinions, findings, and conclusions or recommendations expressed in this material are those of the authors and do not necessarily
reflect the views of the funding sources.

\end{document}